\documentclass[pdftex,twocolumn,epjc3]{svjour3}          

\RequirePackage[T1]{fontenc}

\smartqed  

\RequirePackage{graphicx}
\RequirePackage{mathptmx}      
\RequirePackage{flushend}
\RequirePackage[numbers,sort&compress]{natbib}
\RequirePackage[colorlinks,citecolor=blue,urlcolor=blue,linkcolor=blue]{hyperref}

\usepackage{amsmath}
\usepackage{amssymb}
\usepackage{subfigure}
\usepackage{tabularx}
\usepackage{dcolumn}
\usepackage{bm}
\usepackage{xcolor}
\usepackage[normalem]{ulem}
\usepackage{booktabs,siunitx}
\newcolumntype{C}{>{\centering\arraybackslash}X}

\usepackage{float}

\journalname{Eur. Phys. J. C}

\begin{document}\sloppy

\title{The confining color field in SU(3) gauge theory}

\author{M. Baker\thanksref{e1,addr1}
\and
P. Cea\thanksref{e2,addr2}
\and
V. Chelnokov\thanksref{e3,addr3,addr4}
\and
L. Cosmai\thanksref{e4,addr2}
\and
F. Cuteri\thanksref{e5,addr5}
\and
A. Papa\thanksref{e6,addr3,addr6}
}

\institute{Department of Physics, University of Washington, WA 98105 Seattle, USA\label{addr1}
\and
INFN - Sezione di Bari, I-70126 Bari, Italy\label{addr2}
\and
INFN - Gruppo collegato di Cosenza, I-87036 Arcavacata di Rende, Cosenza, Italy\label{addr3}
\and
\emph{on leave of absence from} Bogolyubov Institute for Theoretical Physics of the National Academy of Sciences of Ukraine\label{addr4}
\and
Institut f\"ur Theoretische Physik, Goethe Universit\"at, 60438 Frankfurt am Main, Germany\label{addr5}
\and
Dipartimento di Fisica dell'Universit\`a della Calabria, I-87036 Arcavacata di Rende, Cosenza, Italy\label{addr6}
}

\thankstext{e1}{e-mail: mbaker4@uw.edu}
\thankstext{e2}{e-mail: paolo.cea@ba.infn.it}
\thankstext{e3}{e-mail: volodymyr.chelnokov@lnf.infn.it}
\thankstext{e4}{e-mail: leonardo.cosmai@ba.infn.it}
\thankstext{e5}{e-mail: cuteri@th.physik.uni-frankfurt.de}
\thankstext{e6}{e-mail: alessandro.papa@fis.unical.it}

\date{Received: date / Accepted: date}

\maketitle

\begin{abstract}
We extend a previous numerical study of SU(3) Yang-Mills theory
in which we measured the 
spatial distribution of all components of the color fields surrounding a static 
quark-antiquark pair for a wide range of quark-antiquark separations, and  
provided evidence that the simulated gauge invariant
 chromoelectric field can be
separated into a Coulomb-like 'perturbative' field and a 'non-perturbative' field,
 identified as the confining part of the SU(3) flux tube field.

In this paper we hypothesize that the fluctuating color fields
 not measured in our simulations
 do not contribute to the string tension. Under this assumption
 the string tension is determined by the color fields 
we  measure, which form
a  tensor $F_{\mu \nu}$ pointing in a single
 direction in color space.  We call this the Maxwell mechanism of confinement.
 
We provide an additional procedure to isolate the non-perturbative (confining)
 field.  We  then extract the string tension  from a stress  energy-momentum tensor 
 $T_{\mu \nu}$ having
   the Maxwell form, constructed from the non-perturbative part of  the tensor
    $F_{\mu \nu}$
   obtained from our simulations. 

To test our hypothesis  we calculate 
 the  string tension from our simulations of the color fields
 for ten values of the quark-antiquark separation ranging from 0.37 fm to 1.2 fm. 
 We  also calculate the spatial distributions of the energy-momentum tensor  $T_{\mu \nu}$ surrounding static quarks
 for this range of separations, and  we compare these distributions with those obtained from direct simulations of
 the energy-momentum tensor in SU(3) Yang-Mills theory.

\end{abstract}

\section{Introduction}

Quantum chromodynamics (QCD) is universally accepted as the theory of 
strong interactions.  Nobody doubts that the well established
phenomenon of confinement of quarks and gluons inside hadrons is encoded
into the QCD Lagrangian. Yet,  our current
understanding does not go beyond that provided by a number of  models
of the QCD vacuum (for a review, see
Refs.~\cite{greensite2011introduction,Diakonov:2009jq}). In particular,
a theoretical {\it a priori} explanation of the so called area law
in large size Wilson loops, which is closely related to a linear
confining potential between a static quark and antiquark at large mutual
distances, is still missing.

In such a challenging situation, first-principle Monte Car\-lo simulations of
QCD on a space-time lattice represent an indispensable tool not only for
checking (or ruling out) models of confinement, but also for providing new
numerical ``phenomenology'' and possibly stimulating original insights into
the mechanism of confinement.

Numerical simulations have established that there is a linear confining
potential between a static quark and antiquark for distances equal to or
larger than about 0.5~fm. This linear
regime extends to infinite distances in SU(3) pure gauge theory, and,
in the presence of dynamical quarks to distances of about 1.4~fm,
where {\em string breaking} should take
place~\cite{Philipsen:1998de,Kratochvila:2002vm,Bali:2005fu}.
The long-distance linear quark-antiquark potential is naturally associated
with a tube-like structure (``flux tube'') of the chromoelectric field
in the longitudinal direction, {\it i.e.} along the line connecting the
static quark and
antiquark~\cite{Bander:1980mu,Greensite:2003bk,Ripka:2005cr,Simonov:2018cbk}.

A wealth of numerical evidence of flux tubes has  accumulated 
 in  SU(2) and SU(3) Yang-Mills theories~\cite{Fukugita:1983du,Kiskis:1984ru,Flower:1985gs,Wosiek:1987kx,DiGiacomo:1989yp,DiGiacomo:1990hc,Cea:1992sd,Matsubara:1993nq,Cea:1994ed,Cea:1995zt,Bali:1994de,Green:1996be,Skala:1996ar,Haymaker:2005py,D'Alessandro:2006ug,Cardaci:2010tb,Cea:2012qw,Cea:2013oba,Cea:2014uja,Cea:2014hma,Cardoso:2013lla,Caselle:2014eka,Cea:2015wjd,Cea:2017ocq,Shuryak:2018ytg,Bonati:2018uwh,Shibata:2019bke}. Most of these
studies concentrated on the shape of the chromoelectric field on the
transverse plane at the midpoint of the line connecting the static quark
and antiquark, given that the other two components of the chromoelectric field and
all the three components of the chromomagnetic field are
suppressed in that plane.

Recent times have witnessed an increasing numerical effort toward a more
comprehensive numerical description of the color field around static sources,
via the measurement of all components of both chromoelectric and
chromomagnetic fields on all transverse planes passing through the  line
between the quarks~\cite{Baker:2018mhw};    of the spatial
distribution of the stress energy momentum tensor~\cite{Yanagihara:2018qqg,Yanagihara:2019foh}; and the
 flux densities for hybrid static potentials~\cite{Bikudo:2018,Mueller:2019mkh}.
A more complete numerical description of the color field around the sources
brings  improved visualization, enabling us to grasp
features otherwise less visible.

In the numerical study~\cite{Baker:2018mhw} we 
simulated the spatial distribution in three dimensions of all  
components of the chromoelectric and chromomagnetic fields generated by 
a static quark-antiquark pair in pure SU(3) lattice gauge theory.  
We found that, although the components of the simulated chromoelectric 
field transverse to the line connecting the pair are smaller than  
the simulated longitudinal chromoelectric field, 
these transverse components are large enough to be fit to a Coulomb-like 
`perturbative' field produced by two static sources parameterized by effective 
charges $\pm Q$ of the sources (see Eq.~\eqref{C2} below).

The longitudinal component of this  Coulomb-like `perturbative' field 
accounts for a fraction of the simulated longitudinal chromoelectric field.
We then identified the remaining longitudinal chromoelectric field as the 
confining `non-perturbative' part of the simulated SU(3) flux tube field.

It is this non-perturbative part of the simulated field which contributes to 
the coefficient of the linear term in the heavy quark potential, 
the string tension.
 
In this paper we extend our simulations to a wider range of 
quark-antiquark separations. 
We extract the string tension from these simulations and compare our analysis 
with the results of recent simulations~\cite{Yanagihara:2018qqg} of the 
energy-momentum tensor in SU(3) Yang-Mills theory.

We present a new procedure (the curl method) to extract a perturbative Coulomb 
field $\vec{E}^{\rm C}$ from the transverse components of the numerically simulated 
chromoelectric field. 
We avoid the use of a fitting function, directly imposing the condition that 
$\vec{E}^{\rm C}$ is irrotational (see Eq.~\eqref{curl_lattice_plaq} below). 
This provides a second method for implementing the underlying idea of our 
previous paper; that is, the chromoelectric field generated by 
a quark-antiquark pair can be separated into perturbative and non-perturbative 
components by a direct analysis of lattice data on the color field 
distributions generated by the pair.
 
As noted in~\cite{Baker:2018mhw}, we can extract the value of the string 
tension from the non-perturbative field by utilizing the fact that the value 
of the chromoelectric field at the position of a quark is the force on 
the quark~\cite{Brambilla:2000gk}.  
However, the Coulomb-like field (Eq.~\eqref{C2}) does not give a good 
description of the transverse components of the chromoelectric field at 
distances closer than approximately two lattice steps from 
the sources~\cite{Baker:2018mhw},  
so that we must use the curl method to isolate the confining field in order 
to extract the string tension directly as the force.

The color fields $F_{\mu \nu}$ we measure, defined by the gauge 
invariant correlation function
$\rho^{\rm conn}_{W, \mu \nu}$ (see Eqs. (\ref{rhoW}) and (\ref{rhoWlimcont}), below),
point in a single direction in color space, parallel to the direction 
of the `source' Wilson loop. In this paper we construct a stress energy -momentum 
tensor $T_{\mu \nu}$  having the Maxwell form from the `measured' flux tube
 field tensor $F_{\mu \nu}$,
and extract the string tension (see~\ref{AppendixStressTensor}). This leads to a picture of a
confining flux tube permeated with lines of force of a gauge invariant 
field tensor $F_{\mu \nu}$ carrying color charge along a single direction.

The `Maxwell' energy-momentum tensor $T_{\mu \nu}$ does not account for the
contribution to the quark-antiquark force from the fluctuating 
color fields not measured in our simulations. On the other hand, the 
complete Yang-Mills energy momentum tensor $T^{\rm YM}_{\mu \nu}$ 
simulated in Ref.~\cite{Yanagihara:2018qqg} includes these fluctuating contributions,
so that comparison  of $T^{\rm YM}_{\mu \nu}$ with the 'Maxwell' 
energy-momentum tensor $T_{\mu \nu}$ constructed from 
the chromoelectric and chromomagnetic fields measured in our simulations provides a measure of the fluctuating contributions to the stress tensor.

We noted in our previous paper that the Coulomb-like `perturbative' field (Eq.~(\ref{C2})) 
generated a stronger long distance Coulomb force between the heavy quarks than the Coulomb force measured in lattice simulations of the heavy quark potential
\cite{Necco:2001xg,Kaczmarek:2005ui,Karbstein:2018mzo}
indicating the  importance of fluctuations for the Coulomb contribution.
In this paper we reexamine this issue.

The paper is organized as follows: in Section~\ref{sec:thBackLatSet} we present the theoretical
background and the lattice setup.  In Section~\ref{sec:distrColFields} we show some results
on the spatial distribution of the color field around the two static sources
and review the procedure to extract its non-perturbative part by subtraction
of the Cou\-lomb-like perturbative part identified by a fit of the transverse
components of the chromoelectric field; in Section~\ref{sect:nonpert} we describe the new
{\em curl} method to isola\-te the non-per\-tur\-bative part; in 
Section~\ref{sec:strTensAndWidth} we
show how to determine the string tension and other fundamental parameters describing the
(non-per\-tur\-bative) flux tube; finally, in Section~\ref{sec:conclAndOutlook} we discuss our results
and give some ideas for future work. 

\section{Theoretical background and lattice setup}\label{sec:thBackLatSet}
  
\begin{figure}[t]
    \centering
    \subfigure[\hspace{2cm}]%
              {\label{fig:operator_Wilson}\includegraphics[width=0.4\textwidth,clip]{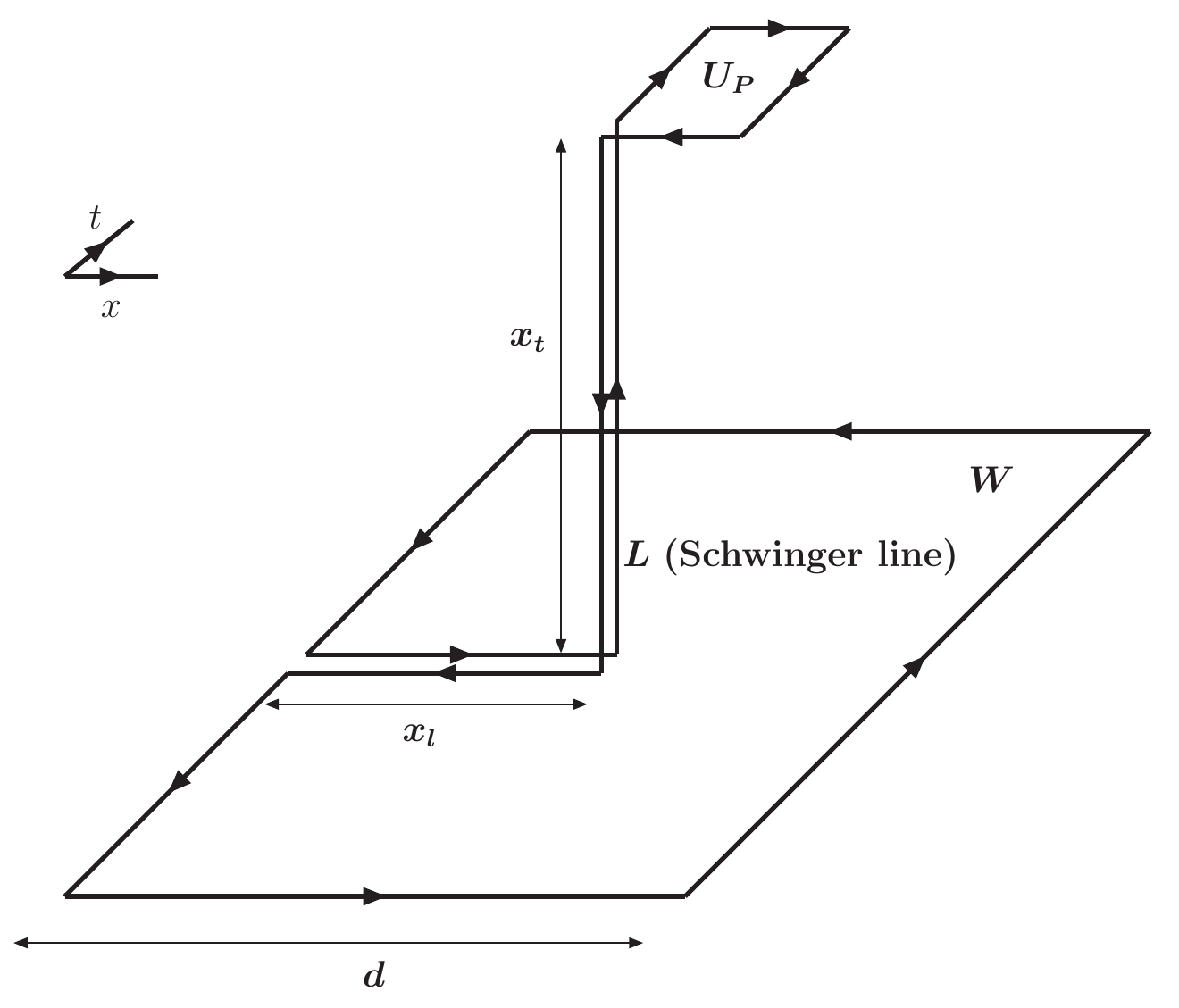}}\hspace{-2cm}
              \subfigure[]%
                        {\label{fig:qqbar}\includegraphics[width=0.15\textwidth,clip]{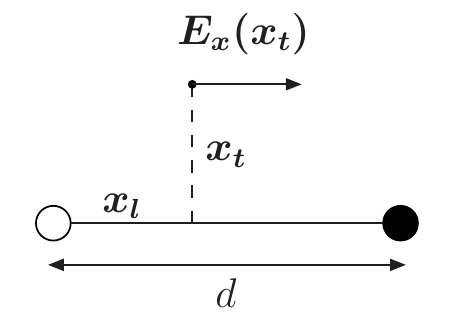}}
                        \caption{~\protect\subref{fig:operator_Wilson} The
                          connected correlator given in Eq.~(\protect\ref{rhoW})
                          between the plaquette $U_{P}$ and the Wilson loop
                          (subtraction in $\rho_{W,\,\mu\nu}^{\rm conn}$ not explicitly drawn).
                          ~\protect\subref{fig:qqbar} The longitudinal
                          chromoelectric field $E_x(x_t)$
                          relative to the
                          position of the static sources (represented by the
                          white and black circles),
                          for a given value of the transverse distance $x_t$.}
                        \label{fig:op_W}
  \end{figure}

  The lattice operator whose vacuum expectation value gives us access to
  the components of the color field generated by a static $q \bar q$ pair is the
  following connected correlator~\cite{DiGiacomo:1989yp,DiGiacomo:1990hc,Kuzmenko:2000bq,DiGiacomo:2000va}:
  \begin{equation}
    \label{rhoW}
    \rho_{W,\,\mu\nu}^{\rm conn} = \frac{\left\langle {\rm tr}
      \left( W L U_P L^{\dagger} \right)  \right\rangle}
        { \left\langle {\rm tr} (W) \right\rangle }
        - \frac{1}{N} \,
        \frac{\left\langle {\rm tr} (U_P) {\rm tr} (W)  \right\rangle}
             { \left\langle {\rm tr} (W) \right\rangle } \; .
  \end{equation}
  Here $U_P=U_{\mu\nu}(x)$ is the plaquette in the $(\mu,\nu)$ plane, connected
  to the Wilson loop $W$ by a Schwinger line $L$, and $N$ is the number of
  colors (see Fig.~\ref{fig:op_W}).

  The correlation function defined in Eq.~(\ref{rhoW}) measures the field 
  strength $F_{\mu\nu}$, since in the naive continuum
  limit~\cite{DiGiacomo:1990hc}
  \begin{equation}
    \label{rhoWlimcont}
    \rho_{W,\,\mu\nu}^{\rm conn}\stackrel{a \rightarrow 0}{\longrightarrow} a^2 g 
    \left[ \left\langle
      F_{\mu\nu}\right\rangle_{q\bar{q}} - \left\langle F_{\mu\nu}
      \right\rangle_0 \right]  \;,
  \end{equation}
  where $\langle\quad\rangle_{q \bar q}$ denotes the average in the presence of 
  a static $q \bar q$ pair, and $\langle\quad\rangle_0$ is the vacuum average.
  This relation is a necessary consequence of the gauge-invariance of the
  operator defined in Eq.~(\ref{rhoW}) and of its linear dependence on the
  color field in the continuum limit (see Ref.~\cite{Cea:2015wjd}).
\begin{figure}[tp]
   \centering
   \subfigure[$E_x(x_t,x_l)$]%
             {\label{fig:FieldsEx}\includegraphics[width=0.45\textwidth,clip]{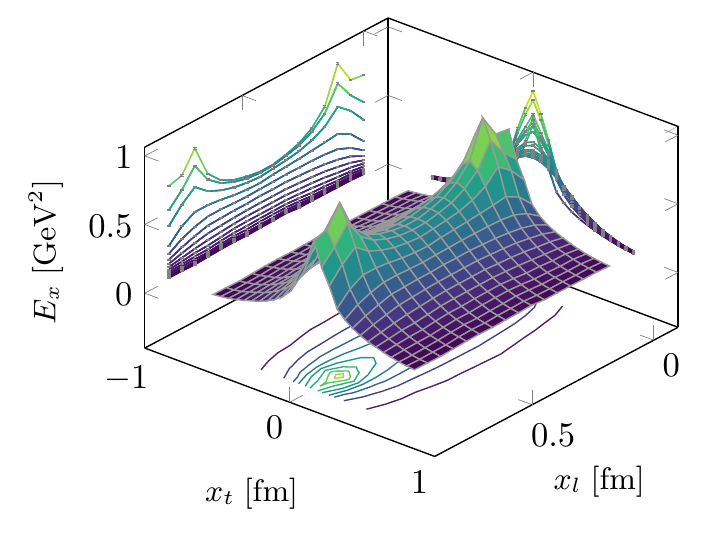}}\vfill
   \subfigure[$E_y(x_t,x_l)$]%
             {\label{fig:FieldsEy}\includegraphics[width=0.45\textwidth,clip]{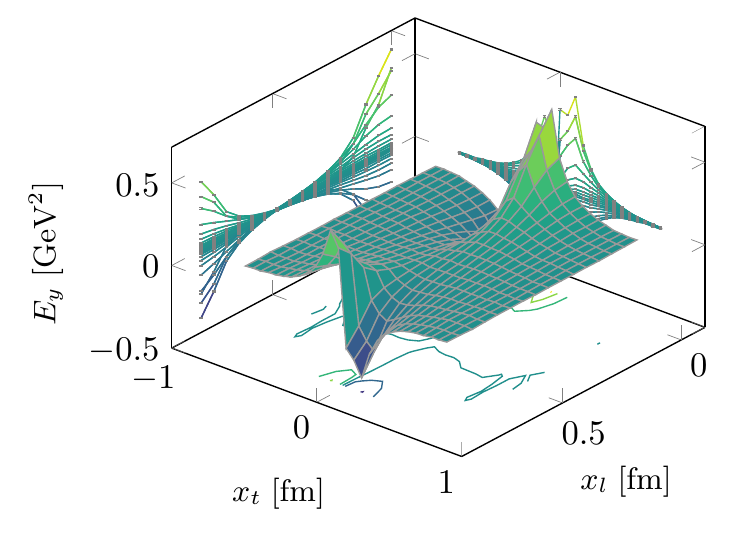}}\vfill
   \subfigure[$E_z(x_t,x_l)$]%
             {\label{fig:FieldsEz}\includegraphics[width=0.45\textwidth,clip]{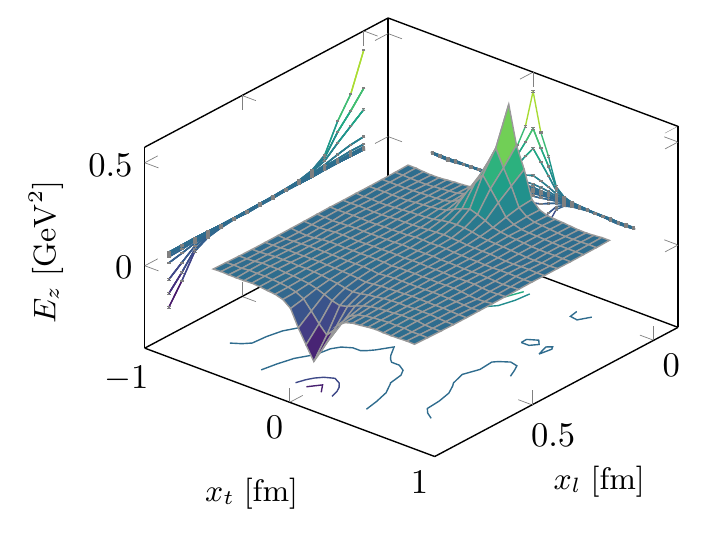}}
   \caption{Surface and contour plots for the three components of the
chromoelectric field at $\beta=6.370$ and $d=0.85$ fm.
All plotted quantities are in physical units.}
   \label{fig:FieldsE}
\end{figure}
\begin{figure}[htb] 
\centering
\includegraphics[width=1.0\linewidth,clip]{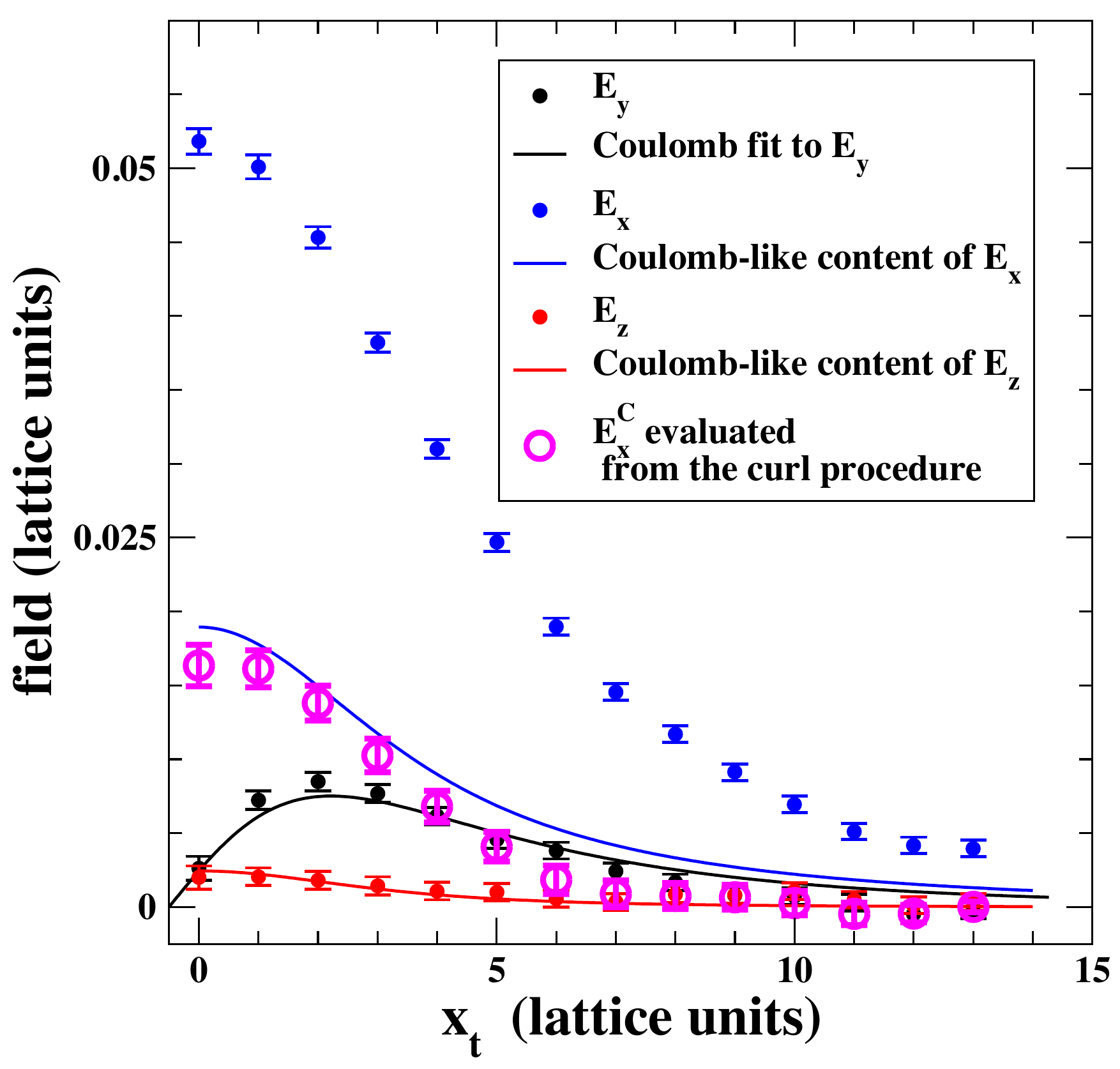}
\caption{The three components of the chromoelectric field measured at
  $\beta=6.240$, $d=16a=1.02\ \mathrm{fm}$, for $x_l = 4a$ 
  and the three components of the perturbative Coulomb field, obtained from
  fitting the transverse $E_y$ field component to the form~(\ref{C2}).}
\label{fig:fields_and_Coulombfit}
\end{figure}
\begin{figure}
   \centering
   \subfigure[$E_x^{\rm NP}(x_t,x_l)$]%
             {\label{fig:Fields_confiningEx}\includegraphics[width=0.45\textwidth,clip]{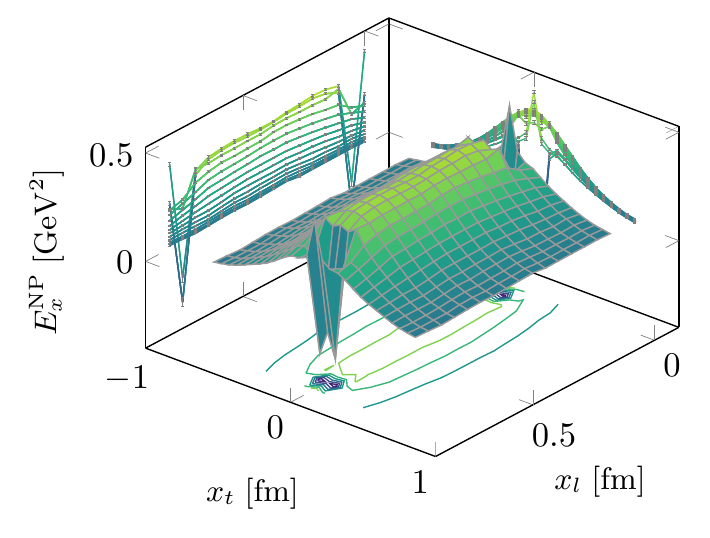}}\vfill
   \subfigure[$E_y^{\rm NP}(x_t,x_l)$]%
             {\label{fig:Fields_confiningEy}\includegraphics[width=0.45\textwidth,clip]{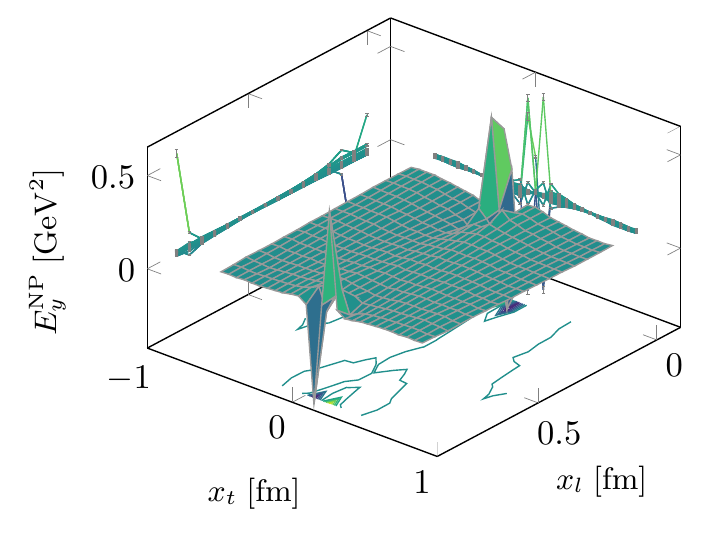}}\vfill
   \subfigure[$E_z^{\rm NP}(x_t,x_l)$]%
             {\label{fig:Fields_confiningEz}\includegraphics[width=0.45\textwidth,clip]{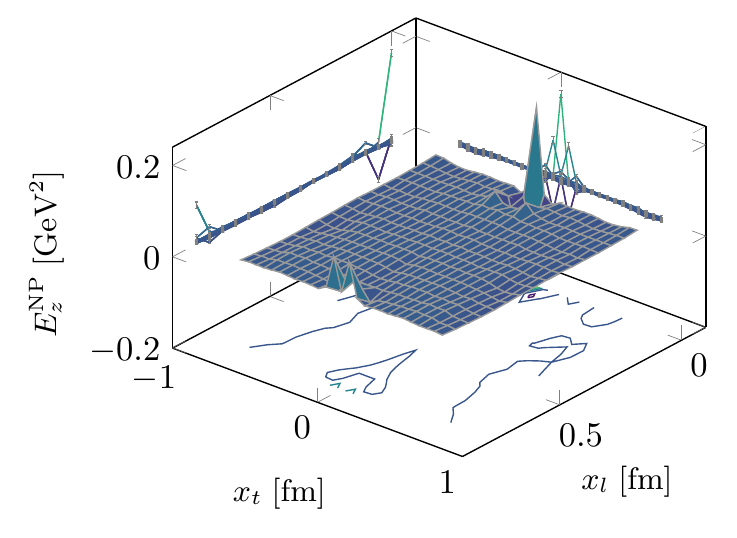}}
\caption{Surface and contour plots for the three components of the non-perturbative
chromoelectric field, $\vec{E}^{\rm NP} \equiv \vec{E} - \vec{E}^{\rm C}$, at $\beta = 6.370$ and
$d = 0.85~$ fm. All plotted quantities are in physical units.}
\label{fig:Fields_confining}
\end{figure}

  The lattice definition of the quark-antiquark field-strength tensor
  $F_{\mu\nu}$ is then obtained by equating the two sides of
  Eq.~(\ref{rhoWlimcont}) for finite lattice spacing.
  In the particular case when the Wilson loop $W$ lies in the plane
  with $\hat \mu=\hat 4$ and $\hat \nu=\hat 1$ (see
  Fig.~\ref{fig:operator_Wilson}) and the plaquette $U_P$ is placed in the
  planes $\hat 4\hat 1$, $\hat 4\hat 2$, $\hat 4\hat 3$, $\hat 2\hat 3$,
  $\hat 3\hat 1$, $\hat 1\hat 2$, we get, respectively, the color field
  components $E_x$, $E_y$, $E_z$, $B_x$, $B_y$, $B_z$, at the spatial point
  corresponding to the position of the center of the plaquette, up to a sign
  depending on the orientation of the plaquette.
  Because of the symmetry  (Fig.~\ref{fig:op_W}), the color fields take on the
  same values at spatial points connected by rotations  around the axis on which
  the sources are located (the $\hat 1$- or $x$-axis in the given example) .
  
  As far as the color structure of the field $F_{\mu \nu}$ is concerned, we note
  that the source of $F_{\mu \nu}$ is the Wilson loop connected to the plaquette
  in Fig.~\ref{fig:op_W}.
  The role of the Schwinger lines entering Eq.~(\ref{rhoW}) is to realize the
  color parallel transport between the source loop and  the ``probe'' plaquette. 
  The Wilson loop defines a direction in color space. The color field $\vec{E}$
  that we measure in  Eq.~(\ref{rhoWlimcont}) points in that direction in the
  color space, {\it i.e.} in the color direction of the source.
  
   There are
  fluctuations of the color fields in the other color directions. 
  We assume that these fluctuating color fields do not contribute to the 
  string tension, so that the flux tube can be described as lines of
  force of the simulated  field $\vec{E}$. 
  
   The simulated flux tube field $\vec{E}$ carries color electric charge
    and color magnetic current along a single direction in color space. 
    The divergence of $\vec{E}$ is equal to the color electric charge 
    density $\rho_{\rm el} (\vec{x})$ and  the 
    curl of $\vec{E}$ is equal to the color magnetic current density
     $\vec{J}_{\rm M} (\vec{x})$.  The  confining force is calculated
     from the divergence of a stress
      tensor $T_{\mu \nu}$  having  the Maxwell form Eq. (\ref{Tmunu})
     .
 
  The operator in Eq.~(\ref{rhoW}) undergoes a non-trivial renormalization,
  which depends on $x_t$, as discussed in a recent work~\cite{Battelli:2019lkz}.
  The procedure outlined in that paper to properly take into account these
  renormalization effects is prohibitively demanding from the computational
  point of view for Wilson loops and Schwinger lines with linear dimension
  of the order of 1~fm, where the interesting physics is expected to take place.
  For this reason, we adopt here the traditional approach to perform smearing
  on the Monte Carlo ensemble configurations before taking measurements
  (see below for details). As shown in the Appendix~A of our previous
  paper~\cite{Baker:2018mhw}, smearing behaves as an effective renormalization,
 effectively pushing the system towards the continuum, where
  renormalization effects become negligible. The {\it a posteriori} validation
  of the smearing procedure is provided by the observation of continuum
  scaling: as carefully checked in Ref.~\cite{Cea:2017ocq}, fields obtained in
  the same {\em physical} setup, but at different values of the coupling, are
  in perfect agreement in the range of parameters used in the present work.
  

  We performed all simulations in SU(3) pure gauge theory, with the
  standard Wilson action as the lattice discretization. A summary of the runs
  performed is given in Table~\ref{tab:runs}. The error analysis was performed
  by the jackknife method over bins at different blocking levels.

%
\begin{table*}[t]
\begin{center} 
  \caption{Summary of the runs performed in the SU(3) pure gauge theory
  (measurements are taken every 100 upgrades of the lattice configuration).}
  \label{tab:runs}
\newcolumntype{Y}{>{\centering\arraybackslash}X}
\begin{tabularx}{0.9\linewidth}{@{}|Y|Y|Y|Y|Y|Y|Y|@{}}
\toprule
$\beta$ & lattice & $a$[fm] & $d$ [lattice] & $d$ [fm] & statistics & smearing steps, $N_{\rm APE}$ \\
\midrule
6.47466 & $36^4$  & 0.047  &  8 &  0.37  &  12900  & 100  \\
6.333   & $48^4$  & 0.056  &  8 &  0.45  &    180  &  80  \\
6.240   & $48^4$  & 0.064  &  8 &  0.51  &   1300  &  60  \\
6.500   & $48^4$  & 0.045  & 12 &  0.54  &   3900  & 100  \\
6.539   & $48^4$  & 0.043  & 16 &  0.69  &   6300  &  100  \\
6.370   & $48^4$  & 0.053  & 16 &  0.85  &   5300  & 100  \\
6.299   & $48^4$  & 0.059  & 16 &  0.94  &  10700  & 100  \\
6.240   & $48^4$  & 0.064  & 16 &  1.02  &  21000  & 100  \\
6.218   & $48^4$  & 0.066  & 16 &  1.06  &  32000  & 100  \\
6.136   & $48^4$  & 0.075  & 16 &  1.19  &  84000  & 120  \\
\bottomrule
\end{tabularx} 
\end{center}
\end{table*}

  We set the physical scale for the lattice spacing according to
  Ref.~\cite{Necco:2001xg}:
\begin{align}
\label{NSscale}
& a(\beta) = r_0 \! \times \! \exp\left[c_0+c_1(\beta\!-\!6)+c_2(\beta\!-\!6)^2+c_3(\beta\!-\!6)^3\right], \nonumber\\
& r_0 = 0.5 \; {\rm fm}, \nonumber\\
& c_0=-1.6804 \,,  c_1=-1.7331 \,,\nonumber\\
& c_2=0.7849 \,, c_3=-0.4428 \,,
\end{align}
for all $\beta$ values in the range $5.7 \le \beta \le 6.92$.
In this scheme the value of the square root of the string tension $\sqrt{\sigma} \approx 0.465 \, {\textrm{GeV}}$
(see Eq. (3.5) in Ref.~\cite{Necco:2001xg}).

The correspondence between $\beta$ and the distance $d$ shown in
Table~\ref{tab:runs} was obtained from this parameterization. 
Note that the distance in lattice units between quark and antiquark, 
corresponding to the spatial size of the Wilson loop in the connected
correlator of Eq.~\eqref{rhoW}, was varied in the range $d=8\,a$ to $d=16\,a$.

  The connected correlator defined in Eq.~(\ref{rhoW}) exhibits large
  fluctuations at the scale of the lattice spacing, which are responsible
  for a bad signal-to-noise ratio. To extract the physical information carried
  by fluctuations at the physical scale (and, therefore, at large distances
  in lattice units) we smoothed out configurations by a {\em smearing}
  procedure. Our setup consisted of (just) one step of HYP
  smearing~\cite{Hasenfratz:2001hp} on the temporal links, with smearing
  parameters $(\alpha_1,\alpha_2,\alpha_3) = (1.0, 0.5, 0.5)$, and
  $N_{\rm APE}$ steps of APE smearing~\cite{Falcioni1985624} on the spatial links,
  with smearing parameter $\alpha_{\rm APE} = 0.25$. 

\section{Spatial distribution of the color fields}\label{sec:distrColFields}

Using Monte Carlo evaluations of the expectation value of the operator
$\rho_{W,\,\mu\nu}^{\rm  }$ over smeared ensembles, we have determined the six
components of the color fields on all two-dimensional planes transverse to the
line joining the color sources allowed by the lattice discretization.
These measurements were carried out for several values of the distance $d$
between the static sources, in the range 0.37~fm to 1.19~fm, at values of
$\beta$ lying inside the continuum scaling region, as determined in
Ref.~\cite{Cea:2017ocq}.

We found that the chromomagnetic field is everywhere much smaller than the
longitudinal chromoelectric field and is compatible with zero within
statistical errors (see, {\it e.g.}, Fig.~3 of Ref.~\cite{Baker:2018mhw}).
As expected, the dominant component of the chromoelectric field is
longitudinal, as is seen in Fig.~\ref{fig:FieldsE}, where we plot the
components of the simulated chromoelectric field $\vec{E}$ at $\beta = 6.370$
as functions of their longitudinal displacement from one of the quarks,
$x_l$, and their transverse distance from the axis, $x_t$.

While the transverse components of the chromoelectric field are also smaller
than the longitudinal component, they are larger than the statistical errors
in a region wide enough that we can match them to the transverse components
of an effective Coulomb-like field $\vec{E}^{\rm C}$ produced by two
static sources. 
For points which are not very close to the quarks, this matching can be
carried out with a single fitting parameter $Q$, the effective charge of 
static quark and antiquark sources determining $\vec{E}^{\rm C}$. %

To the extent that we can fit the transverse components of the simulated field
$\vec{E}$ to those of $\vec{E}^{\rm C}$ with an appropriate choice of $Q$,
the non-perturbative difference $\vec{E}^{\rm NP}$ between the simulated
chromoelectric field $\vec{E}$ and the effective Coulomb field $\vec{E}^{\rm C}$,
\begin {equation}
\vec{E}^{\rm NP} ~\equiv~\vec{E}~-~\vec{E}^{\rm C}\;,  
\label{ENP}
\end{equation}
will be purely  longitudinal. We then identify $\vec{E}^{\rm NP}$ as the confining
field of the QCD flux tube.

To illustrate this idea, let us fix, for the sake of definiteness,
$\beta=6.240$ and put the two sources at a distance $d=16a=1.02~$ fm.
We then consider the plane, transverse to the longitudinal $x$-axis
connecting the two sources, at a distance $x_l = 4a$ from one of them, and evaluate the
components $E_x$, $E_y$ and $E_z$ of the chromoelectric field in this
transverse plane. The lattice determinations of $E_y$ on this plane can be
fitted by the $y$-component of an effective Coulomb field
%
\begin{eqnarray}
\label{C2}
& \vec{E}^{\rm C} (\vec{r}) \; = \;  Q \left( \frac{\vec{r}_1}{\max(r_1, R_0)^3} \;
- \;  \frac{\vec{r}_2}{\max(r_2, R_0)^3} \; \right ) \; ,
\\ \nonumber
& \vec{r}_1\equiv \vec r -\vec r_Q\;, \;\;\;
\vec{r}_2\equiv\vec r - \vec r_{-Q}\;,
%
\end{eqnarray}
where $\vec r_Q$ and $\vec r_{-Q}$ are the positions of the two static color
sources and $R_0$ is the effective radius of the color source, introduced to
explain, at least partially, the decrease of the field close to the sources.
This fit is shown in Fig.~\ref{fig:fields_and_Coulombfit} -- see black dots
and black solid line. Using the values of the fit parameters $Q$ and $R_0$
obtained by the fit of $E_y$, one can construct $E_z^{\rm C}$ and
$E_x^{\rm C}$ and compare them to lattice data. Furthermore,  the
Coulomb-like content of $E_z$ fully accounts for  the
$z$-component of the chromoelectric field (see red dots and red solid line in
Fig.~(\ref{fig:fields_and_Coulombfit})), but $E_x^{\rm C}$ accounts for only
a fraction of the  longitudinal component of the chromoelectric field
(see blue dots and blue solid line in Fig.~(\ref{fig:fields_and_Coulombfit})).
This strongly suggests that the non-perturbative component of the chromoelectric
field is almost completely oriented along the longitudinal direction. It
can be isolated once the parameters of the Coulomb-like component are
determined by a fit to the $y$- and/or $z$-components of the lattice
determination of the chromoelectric field.

The procedure we have just illustrated in a specific case, can be carried out
in a systematic manner. We observe that in making the fit we must take into
account that the color fields are probed by a plaquette, so that the measured
field value should be assigned to the center of the plaquette. This also means
that the $z$-component of the field is  probed at a distance of $1/2$ lattice
spacing from the $x\,y$ plane, where the $z$-component  of the Coulomb field
$E_z^{\rm C}$ is non-zero and can be matched with the measured value $E_z$ for the
same value of $Q$. For further details about the fitting procedure
and the extraction of the fit parameters we refer to  Appendix~B of
Ref.~\cite{Baker:2018mhw}.

In Table~\ref{tab:coulombFit}, we list the values of the effective charge
$Q$ obtained from  the lattice measurements of $E_z$ and $E_y$ at the values of
$d$, the quark-antiquark separation, considered in this work.
%
\begin{table*}[t]
\begin{center}
  \caption{Values of the fit parameters $Q$ and $R_0$ extracted from Coulomb
    fits of the transverse components of the chromoelectric field and values
    of the longitudinal chromoelectric fields at $(d/2,0)$, the midpoint 
    between the sources and transverse distance zero, for several values of
    distance $d$. $E_x(d/2,0)$ is the unsubtracted simulated field and
    $E_{x}^{\rm NP}(d/2,0)$ is the non-perturbative chromoelectric field.
    For comparison, in the last column the non-perturbative chromoelectric field
    ${E_{x}^{\rm NP}}_{\mathrm{curl}}(d/2,0)$ obtained using the irrotational property of the perturbative field (see section~\ref{sect:nonpert}) is given.
    For the parameters of the Coulomb fit we quote, along with the statistical
    error, a systematic uncertainty that accounts for the variability in the
    values of the fit parameters extracted from all acceptable fits to
    $E_{y}$ and $E_{z}$ at different $x_l$ values (for more details, see
    Appendix~B of Ref.~\cite{Baker:2018mhw}). As the distance between the sources is made smaller and smaller the quality of the Coulomb fits deteriorates and $Q$ and $R_0$ cannot be reliably extracted for $d\le0.51$ fm. }
\label{tab:coulombFit}
\begin{tabularx}{0.775\linewidth}{@{}|S[table-format=1.5]|S[table-format=1.2]|S[table-format=1.12]|S[table-format=1.12]|S[table-format=1.8]|S[table-format=1.8]|S[table-format=1.9]|@{}}
\toprule
 {$\beta$} & {$d$ [fm]} & {$Q$} & {$R_0$ [fm]} & {$E_x(d/2,0)$} & {$E_{x}^{\rm NP}(d/2,0)$} & {${E_{x}^{\rm NP}}_{\mathrm{curl}}(d/2,0)$}\\
  &  &  &  & {$[\mathrm{GeV}^2]$} & {$[\text{GeV}^2]$} & {$[\text{GeV}^2]$}\\
\midrule
6.47466   & 0.37 & {-}  & {-}  & 1.00155(22)  & {-}  & 0.33581(20) \\
6.333     & 0.45 & {-}  & {-}  & 0.8086(7)  & {-} & 0.3388(9) \\
6.240     & 0.51 & {-}  & {-}  & 0.7059(3)  & {-} & 0.35353(29) \\
6.500     & 0.54 & 0.2736(13)(875)  & {-}  & 0.6550(5)  &  0.35762(18) & 0.3584(6) \\
6.539     & 0.69 & 0.2729(4)(16)  & {-}  & 0.5204(16) & 0.3378(5) & 0.3683(25) \\
6.370     & 0.85 & 0.262(3)(131)  &  0.0975(6)(60) & 0.446(4)   & 0.3331(16) & 0.348(7) \\
6.299     & 0.94 & 0.259(5)(31)  & 0.1112(10)(205)  & 0.424(6)   & 0.332(3) & 0.325(8)  \\
6.240     & 1.02 & 0.2877(10)(108)  &  0.1183(21)(287) & 0.418(8)   & 0.331(5) & 0.343(12) \\
6.218     & 1.06 & 0.293(6)(169)  & 0.1211(22)(359)  & 0.398(9)   & 0.315(6) & 0.347(13) \\
6.136     & 1.19 & 0.314(8)(89)  &  0.160(5)(28) & 0.359(29)  & 0.29(3) & 0.33(4) \\
\bottomrule
\end{tabularx}
\end{center}
\end{table*}
The statistical uncertainties in the quoted $Q$ values result from the
comparisons among Coulomb fits of $E_y$ and $E_z$ at the values of $x_l$, 
for which we were able to get meaningful results for the fit. 
The values of $R_0$ in physical units grow with the lattice step $a$, while in lattice units they show more stability.
This suggests that the effective size of a color charge in our case 
is mainly explained by lattice discretization artifacts and the smearing
procedure, and is not a physical quantity (see Appendix~B of
Ref.~\cite{Baker:2018mhw}). 

Evaluating the contribution of the field of the quark to $\bold{E}^{\rm C}$
in Eq.~(\ref{C2}) at the position $\bold{r_{-Q}}$ of the antiquark and
multiplying by the charge $-4 \pi Q$ of the antiquark yields a Coulomb force
between the quark and antiquark with coefficient $- 4 \pi Q^2$.  By comparison,
the standard string picture of the color flux tube gives a Coulomb correction
of strength $-\pi/12$ to the long distance linear potential  (the universal
L\"uscher term arising from the long wave length transverse fluctuations of
the flux tube~\cite{Luscher:1980ac}). In addition, the strength $\frac{\pi}{12}$
 of the Luscher term is approximately equal to the strength of the Coulomb force 
 determined from the analysis of lattice simulations of the heavy quark
 potential at distances down to about $0.4$ fm.  \cite {Necco:2001xg, Karbstein:2018mzo}.
 
 By contrast, the strength $-4 \pi  Q^2$ of the Coulomb force generated by $\vec{E}^{\rm C}$ is roughly 4 times larger than $\frac{\pi}{12}$ for the values of the effective charge $Q$ listed in Table (\ref{tab:coulombFit})
 and determined from our simulations of $\rho^{\rm conn}_{W, \mu, \nu}$. 
 Therefore the fluctuating color fields not measured in our simulations  must
 be taken into account in calculating the Coulomb correction to the long distance heavy 
 quark potential.

\begin{figure}[htb] 
\centering
\includegraphics[width=1.0\linewidth,clip]{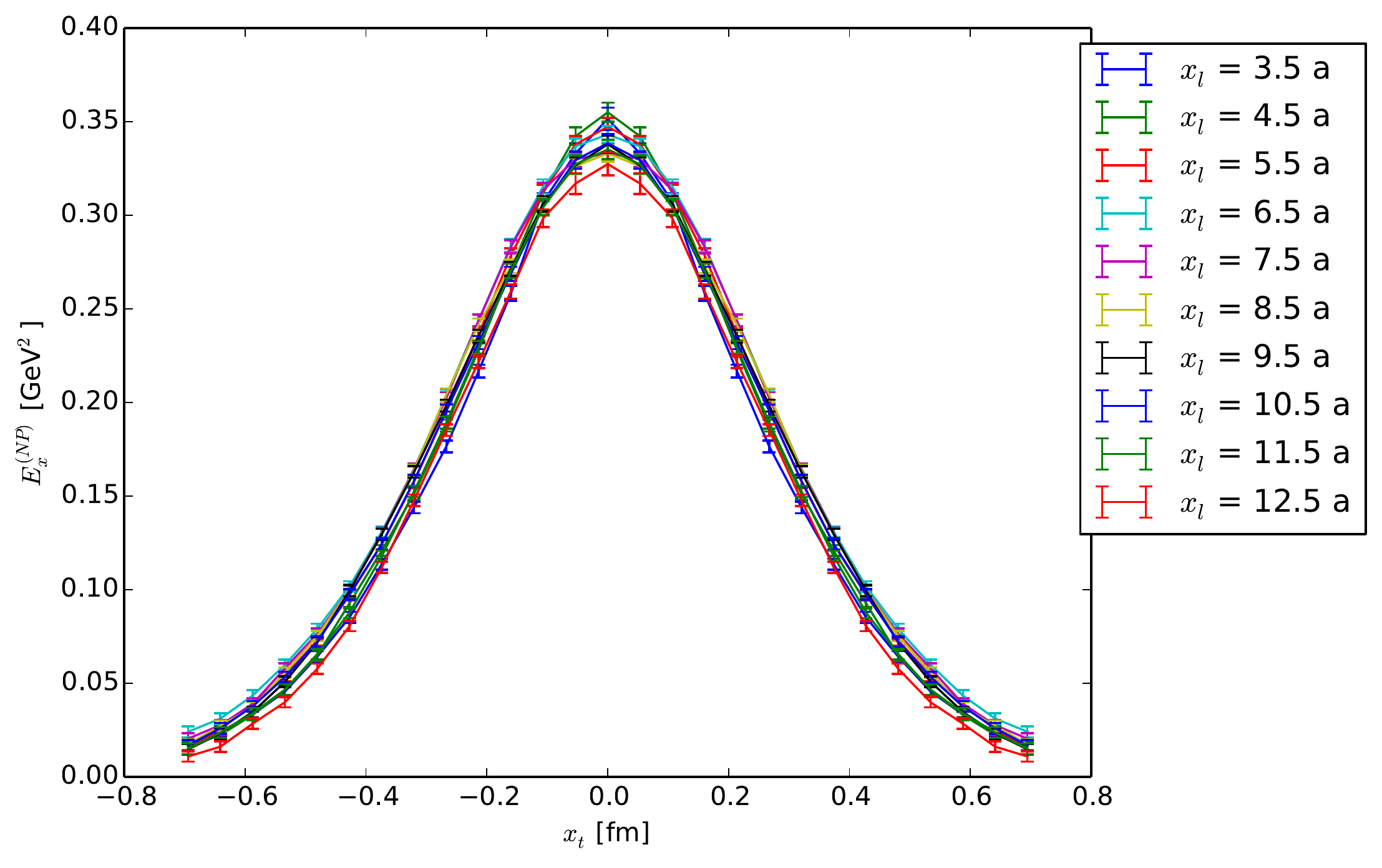}
\caption{Transverse cross sections of the non-perturbative field
$E_x^{\rm NP}(x_t)$ at $\beta=6.370$, $d=0.85$ fm, for several values of $x_l$.}
\label{fig:NPEx}
\end{figure}

In Fig.~\ref{fig:Fields_confining} we plot the longitudinal component $E_x^{\rm NP}$
of the non-perturbative field in Eq.~\eqref{ENP} as a function of the longitudinal and
transverse displacements $x_l$, $x_t$ at $\beta = 6.370$. As expected,
$E^{\rm NP}_x$ is almost uniform along the flux tube at distances not too close to
the static color sources. This feature is better seen in Fig.~\ref{fig:NPEx},
where transverse cross sections of the  field  $E_x^{\rm NP} ( x_l, x_t)$, plotted in
Fig.~\ref{fig:Fields_confining}, are shown for the values of $x_l$ specified in
Fig.~\ref{fig:NPEx}. For these values of $x_l$  the shape of the
non-perturbative longitudinal field is basically constant all along the axis. 
A similar scenario holds in the other lattice setups listed in
Table~\ref{tab:runs}.

In Table~\ref{tab:coulombFit} we also compare the values of the measured
longitudinal chromoelectric field $E_x$ with those of  the non-perturbative field
$E_x^{\rm NP}$ on the axis at the midpoint between the quark and antiquark,
for all ten values of their separation $d$. Given that $E^{\rm NP}_x$ is almost
uniform along the axis, $E^{\rm NP}_x =  E_x^{\rm NP} (x_l, x_t=0)$ 
at all points $x_l$ on the axis for all distances larger than approximately
$0.1-0.2~$ fm from the quark sources. 
\section{Non-perturbative content of the flux tube: the {\em curl} method}
\label{sect:nonpert}

  While the Coulomb field~(\ref{C2}) gives a  good description of the transverse 
  components of the chromoelectric field when the distance from the sources is 
  not too small, it does not give a good description at smaller distances, approximately 
  two lattice spacings from the sources.
  This can be either the result of the non-spherical form of the effective charges,
  or an effect introduced by the discrete lattice.

  To extract the confining  part of the chromoelectric field in the data it is then
preferable to have a procedure which avoids the use of an explicit fitting function, and which can work close to the quark sources. With this aim in mind we use the following two steps to separate the field into 'perturbative'
and 'non perturbative' components.

\begin {enumerate}
\item We identify the transverse component $E_y$ of the simulated field with the transverse component $E^{\rm C}_y$ of the perturbative field, $E^{\rm C}_y \equiv E_y$.

\item We impose the condition that the perturbative field is irrotational, curl $\vec{E}^{\rm C} =0$.
\end{enumerate}

Condition (1) implies that the nonperturbative field is purely longitudinal, ${E}^{\rm NP}_y =0$.
  Condition (2) will then fix the longitudinal component  ${E}^{\rm C}_x$ of the perturbative field as well as the longitudinal component $E^{\rm NP}_x = E_x - E^{\rm C}_x$ of the non-perturbative field.
   
To implement the irrotational condition (2), taking into account that the fields 
  are measured at discrete lattice points, the sum of the measured fields
  along any closed lattice path is zero. For example, on a plaquette this
  amounts to 
  \begin{equation}
  \label{curl_lattice_plaq}
    E_x^{\rm C} (x,y) + E_y^{\rm C} (x+1, y) - E_x^{\rm C} (x,y+1) - E_y^{\rm C} (x, y) = 0\ .
  \end{equation}
One can easily solve this equation for $E_x^{\rm C}$ obtaining
  \begin{equation}
  \label{curl_Ex}
   E_x^{\rm C} (x,y) = \sum_{y^\prime = y}^{y_{\mathrm{max}}} 
    \left(E_y (x, y^\prime) - E_y (x + 1, y^\prime)\right) 
    + E_x^{\rm C} (x, y_{\mathrm{max}} + 1)\ .
  \end{equation}
  This of course leaves one unknown on each transverse slice of the field -- 
  the value of $E_x^{\rm C} (x, y_{\mathrm{max}} + 1)$, but if the value of $y_{\max}$ 
  is large enough, the perturbative field at that distance should already be
  small, so in our analysis we just put $E_x^{\rm C} (x, y_{\mathrm{max}} + 1) = 0$.
  To check that this indeed makes a little change to our results, we have used 
  a separate procedure in which we fixed $E_x^{\rm C} (x, y_{\mathrm{max}} + 1) = E_x$, 
  in practice making $E_x^{\rm NP} = 0$ at  the largest transverse distance. 
This procedure gave similar results. 

\begin{figure}
   \centering
   \subfigure[$E_x(x_t,x_l)$]%
             {\label{fig:Field_full_Ex}\includegraphics[width=0.45\textwidth,clip]{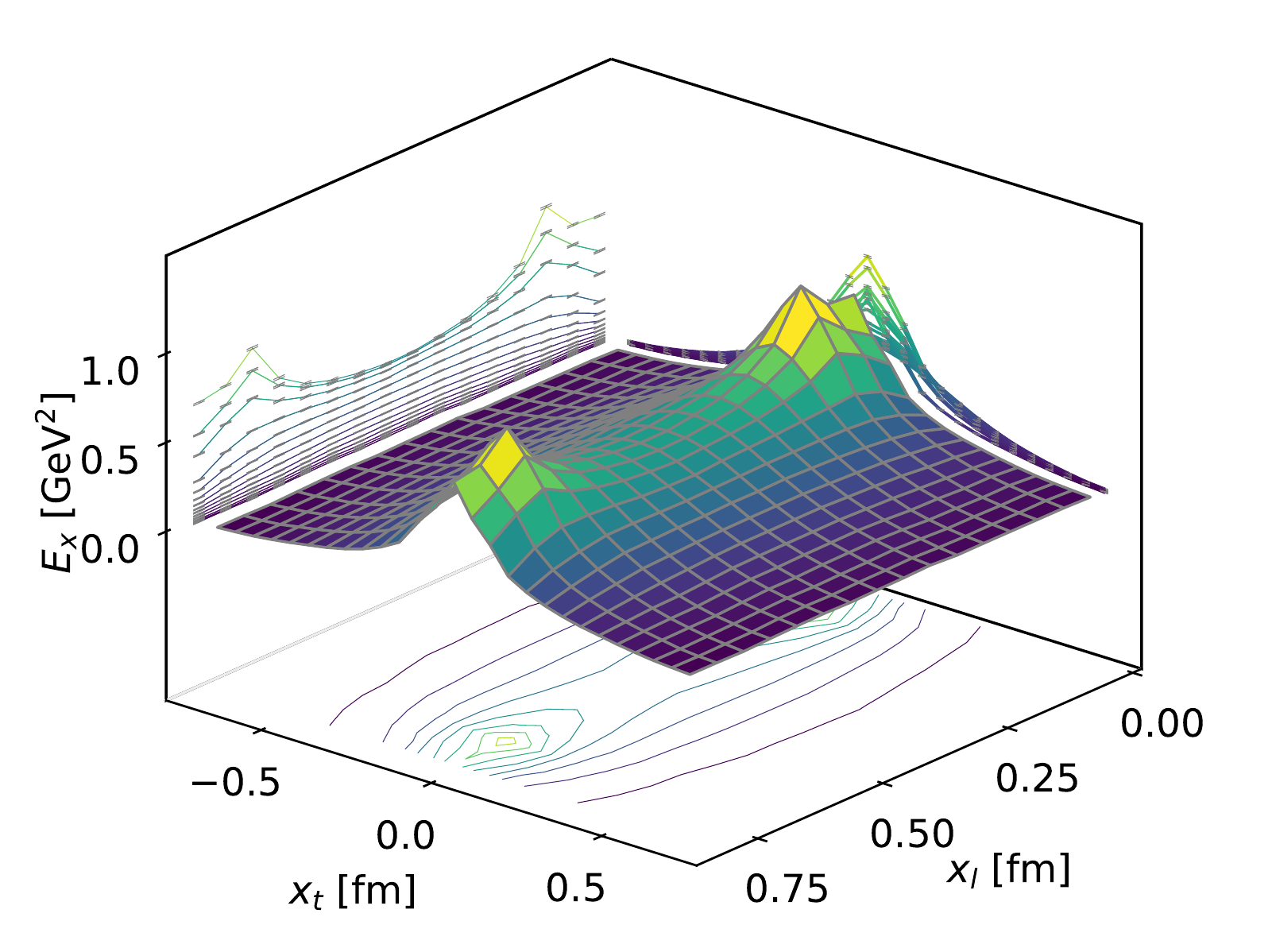}}\vfill
   \subfigure[$E_x^{\rm C}(x_t,x_l)$]%
             {\label{fig:Field_perturbative_Ex}\includegraphics[width=0.45\textwidth,clip]{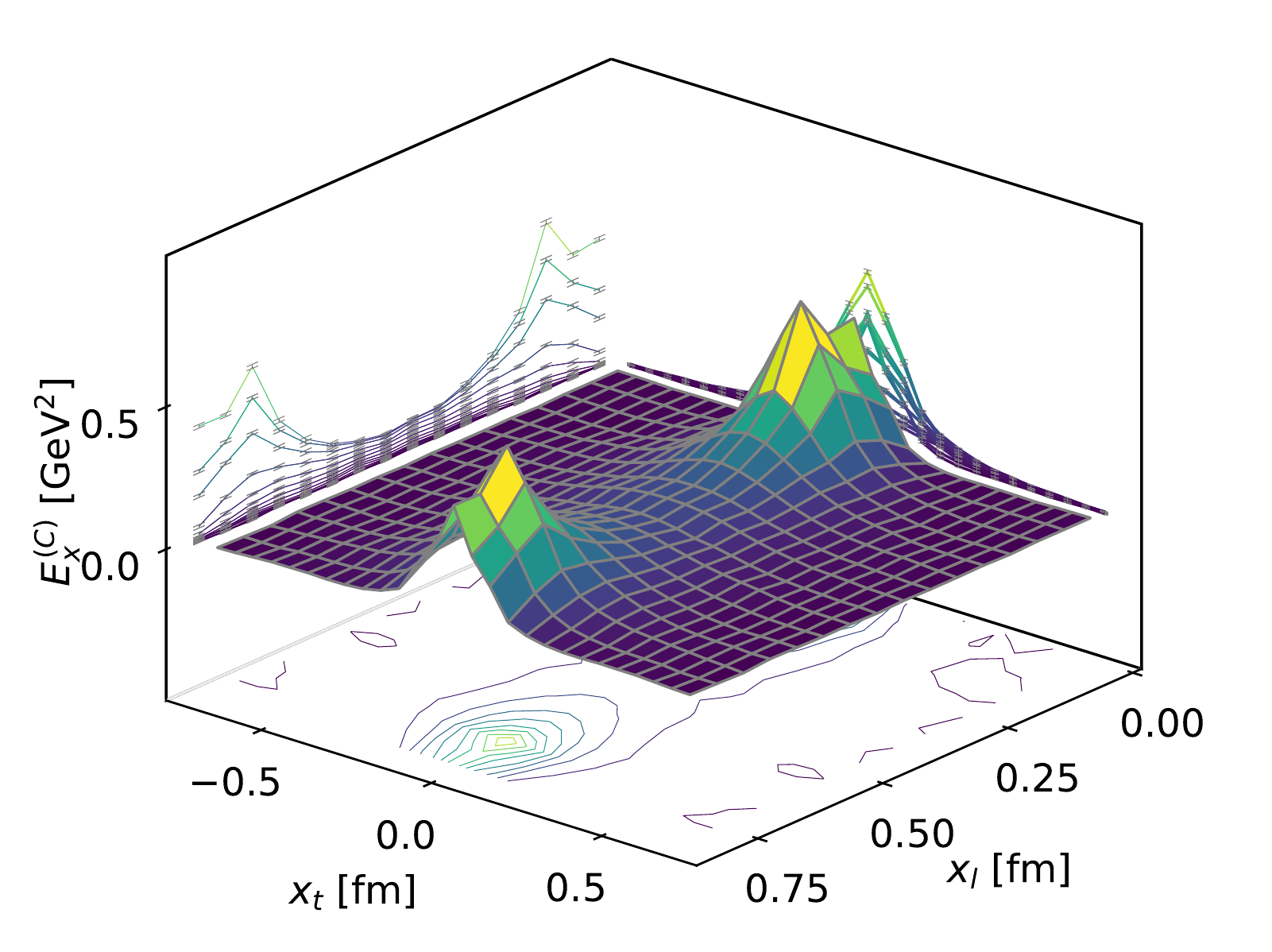}}\vfill
   \subfigure[$E_x^{\rm NP}(x_t,x_l)$]%
             {\label{fig:Field_nonperturbative_Ex}\includegraphics[width=0.45\textwidth,clip]{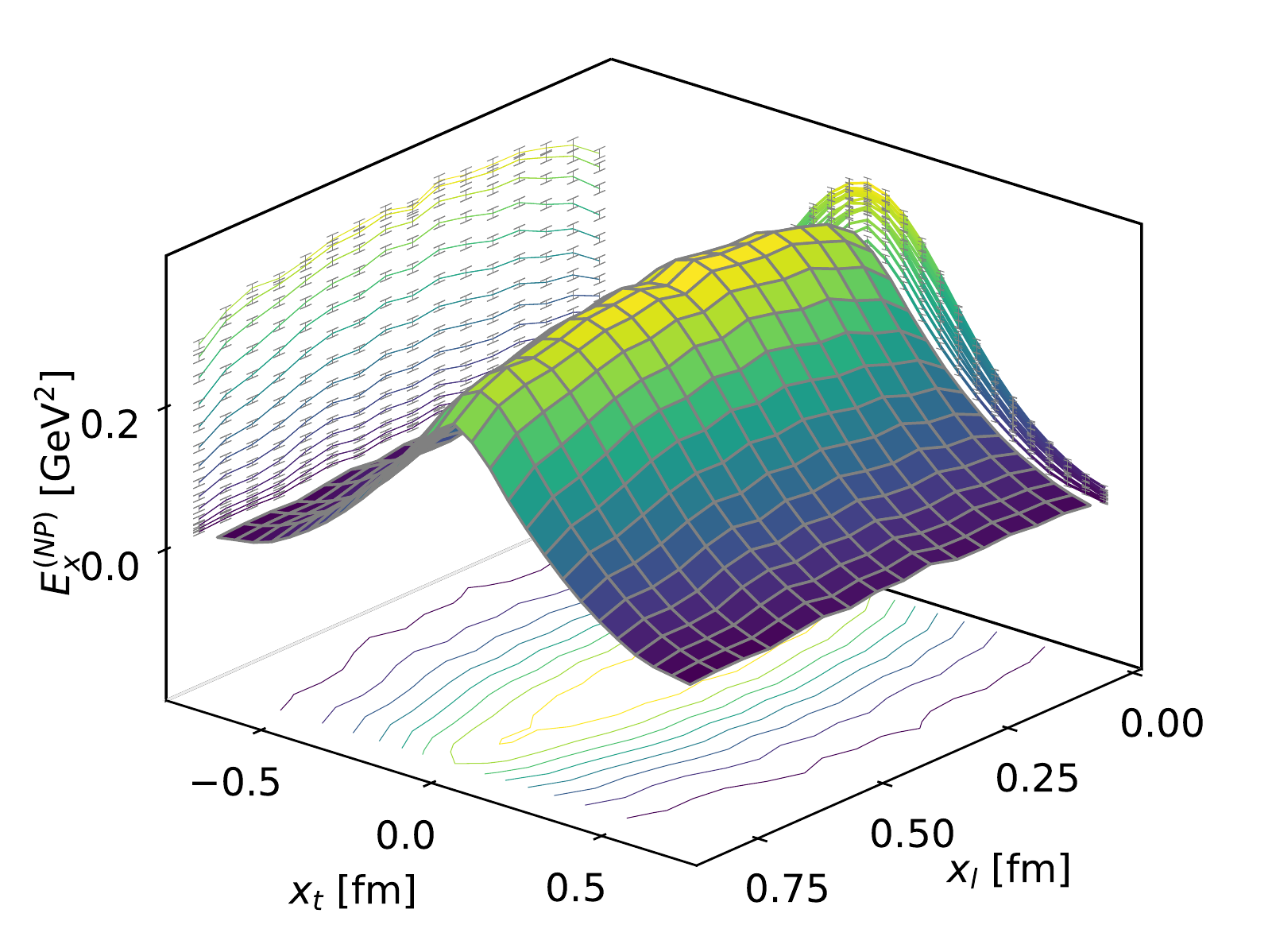}}
\caption{Surface and contour plots for the longitudinal components of the full, perturbative and non-perturbative 
chromoelectric field obtained by using the curl procedure at $\beta = 6.370$ and
$d = 0.85~ $fm. All plotted quantities are in physical units.}
\label{fig:curl_subtraction_surfaces}
\end{figure}

\begin{figure}[htb] 
\centering
\includegraphics[width=1.0\linewidth,clip]{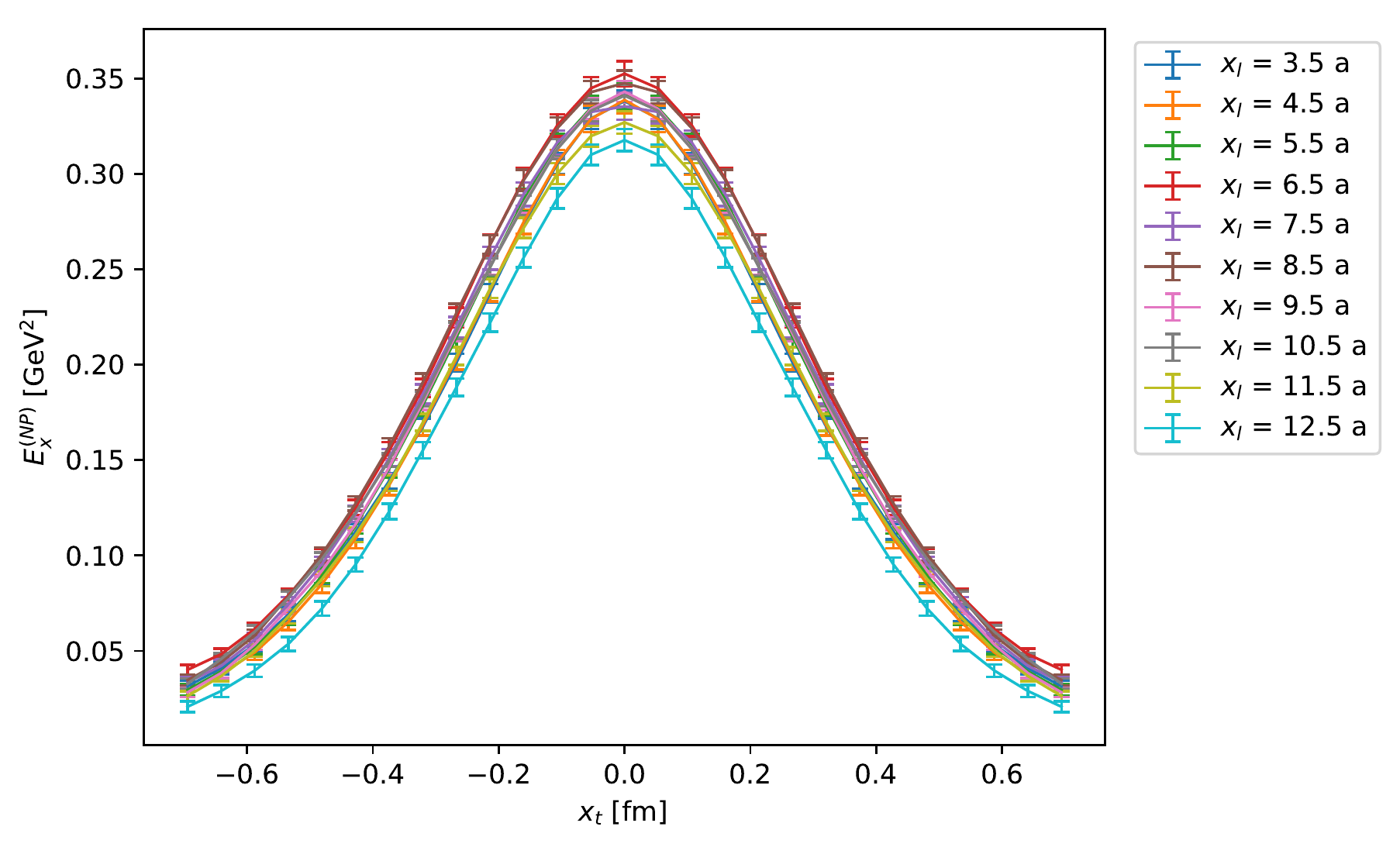}
\caption{Transverse cross sections of the non-perturbative field obtained by using the curl procedure
$E_x^{\rm NP}(x_t)$ at $\beta=6.370$, $d=0.85$ fm, for several values of $x_l$.
}
\label{fig:curl_NPEx}
\end{figure}

  After the estimation of the perturbative longitudinal field $E_x^{\rm C}$ one can
  subtract it from the total field, obtaining the non-perturbative component 
  (see Fig.~\ref{fig:curl_subtraction_surfaces}). One can see that 
  the non-perturbative part of the flux tube exhibits very little change along 
  the line connecting the quark-antiquark pair; 
 even at the smallest distances
from the sources the non-perturbative field remains smooth  (This is seen more
clearly in Fig.~\ref{fig:curl_NPEx}).

\begin{table*}[t]
\begin{center} 
  \caption{The Clem parameters describing the non-perturbative field transverse
    section going through the midpoint between the quark and antiquark positions. 
    The data for the fit is obtained using the curl subtraction method, taking the 
    perturbative field at $y_{\mathrm{max} + 1}$ equal to zero.}
  \label{tab:clem_parameters}
\newcolumntype{Y}{>{\centering\arraybackslash}X}
\begin{tabularx}{0.628\textwidth}{@{} |S[table-format=1.5]|S[table-format=1.2]|S[table-format=1.7]|S[table-format=2.7]|S[table-format=1.7]|S[table-format=1.6]|S[table-format=3.2]| @{}}
\toprule
{$\beta$} & {$d$ [fm]} & {$\varphi$} & {$\mu$ [fm${}^{-1}$]} & {$\alpha$} & {$\kappa$} & {$\chi^2_r$} \\ 
\midrule
6.47466 & 0.37     &  3.474(4)  & 4.999(9)  & 1.192(4) & 0.808(4) & 318  \\
6.333   & 0.45     &  3.83(3)   & 5.30(6)   & 1.55(3) & 0.576(15)  & 12.0   \\
6.240   & 0.51     &  4.028(11) & 6.039(26) & 2.141(20) & 0.375(5) & 43.5   \\
6.500   & 0.54     &  4.370(15) & 5.71(4)   & 2.02(3) & 0.406(9)  & 4.46 \\
6.539   & 0.69     &  4.50(7)   & 6.25(20)  & 2.47(16) & 0.309(27)  & 0.03  \\
6.370   & 0.85     &  5.40(25)  & 6.7(9)    & \multicolumn{1}{S[table-format=1.1,table-number-alignment = left]|}{4.0{(1.1)}}  & 0.17(7) & 0.06  \\
6.299   & 0.94     &  5.2(4)    & \multicolumn{1}{S[table-format=2.1,table-number-alignment = left]|}{7.8{(1.9)}}  & \multicolumn{1}{S[table-format=1.1,table-number-alignment = left]|}{5.5{(2.8)}} & 0.10(7)  & 0.02  \\
6.240   & 1.02     &  8.0(7)    & 4.4(8)    & 2.4(9) &  0.33(17)  & 0.18  \\
6.218   & 1.06     &  6.6(7)    & \multicolumn{1}{S[table-format=2.1,table-number-alignment = left]|}{6.0{(1.8)}}  & \multicolumn{1}{S[table-format=1.1,table-number-alignment = left]|}{4.0{(2.4)}} &  0.16(13)  & 0.05  \\
6.136   & 1.19     &  \multicolumn{1}{S[table-format=1.1,table-number-alignment = left]|}{5.5{(1.6)}}  & \multicolumn{1}{S[table-format=2.0]|}{81(27)}    & {7(5)${}\times 10^2$} & {8(9)${}\times 10^{-5}$} & 0.17  \\
\bottomrule
\end{tabularx}
\end{center}
\end{table*}

\begin{table*}[t]
\begin{center} 
  \caption{The string tension estimated using the non-perturbative field
    from the curl procedure by employing
    different methods (from left to right: numerical integration of the field,
    analytical integration of the Clem function with parameters given in 
    Table~\ref{tab:clem_parameters}, estimation of fields at sources). In the
    last column we report also the value of the string tension obtained
    by numerically integrating Eq.~(\ref{sqrtstring-clem}) and using the non-perturbative
    field from the Coulomb subtraction $E_x^{\rm Coulomb}$.}
  \label{tab:string_tension}
\begin{tabularx}{0.66\textwidth}{@{} |S[table-format=1.5]|S[table-format=1.2]|S[table-format=1.7]|S[table-format=1.7]|S[table-format=2.9]|S[table-format=1.7]| @{}}
\toprule
$\beta$ & {$d$ [fm]} & {$\sqrt{\sigma_{\rm int}}$ [GeV]} & {$\sqrt{\sigma_{\rm Clem}}$ [GeV]}& {$\sqrt{\sigma_{\rm 0}}$ [GeV]}& {$\sqrt{\sigma_{\rm Coulomb}}$ [GeV]}\\ 
\midrule
6.47466 & 0.37     &  0.4591(3)   & 0.4659(3) & 0.53426(22)   & {-}           \\
6.333   & 0.45     &  0.5020(19)  & 0.5045(20)  & 0.5313(6)   & {-}           \\
6.240   & 0.51     &  0.5409(10)  & 0.5430(10)  & 0.5340(4)   & {-}           \\
6.500   & 0.54     &  0.5582(9)   & 0.5687(10)  & 0.5410(7)   & 0.491 (25)  \\
6.539   & 0.69     &  0.583(4)    & 0.596(5)    & 0.5526(28)  & 0.468 (4)   \\
6.370   & 0.85     &  0.633(16)   & 0.640(17)   & 0.528(7)    & 0.412 (17)  \\
6.299   & 0.94     &  0.617(23)   & 0.620(24)   & 0.527(11)   & 0.598 (7)   \\
6.240   & 1.02     &  0.75(4)     & 0.77(4)     & 0.520(17)   & 0.616 (7)   \\
6.218   & 1.06     &  0.69(4)     & 0.62(3)     & 0.482(19)   & 0.599 (24)  \\
6.136   & 1.19     &  0.67(11)    & 0.67(12)    & 0.56(5)     & 0.593 (28)  \\
\bottomrule
\end{tabularx}
\end{center}
\end{table*}

\begin{table*}[t]
\begin{center} 
  \caption{The flux tube width estimated using the non-perturbative field
    from the curl procedure by employing 
    different methods (from left to right: numerical integration of the field,
    analytical integration of the Clem function with parameters given in 
    Table~\ref{tab:clem_parameters}). In the last column we report also the
    value of the width obtained by numerically integrating
    Eq.~(\ref{width}) and using the non-perturbative field from the Coulomb
    subtraction
    $E_x^{\rm Coulomb}$.}
  \label{tab:width}
\begin{tabularx}{0.52\textwidth}{@{} |S[table-format=1.5]|S[table-format=1.2]|S[table-format=1.9]|S[table-format=1.7]|S[table-format=1.8]| @{}}
\toprule
$\beta$ & {$d$ [fm]} & {$\sqrt{w^2_{\rm int}}$ [fm]} & {$\sqrt{w^2_{\rm Clem}}$ [fm]} & {$\sqrt{w^2_{\rm Coulomb}}$ [fm]}\\ 
\midrule
6.47466 & 0.37     &  0.31696(6)  & 0.4795(6)    & {-}           \\
6.333   & 0.45     &  0.3598(7)   & 0.477(3)     & {-}           \\
6.240   & 0.51     &  0.3838(3)   & 0.4543(9)    & {-}           \\
6.500   & 0.54     &  0.31716(15) & 0.4727(18)   & 0.313(11)  \\
6.539   & 0.69     &  0.3061(5)   & 0.457(6)     & 0.3020(23) \\
6.370   & 0.85     &  0.3712(24)  & 0.497(21)    & 0.343(16)  \\
6.299   & 0.94     &  0.393(5)    & 0.483(29)    & 0.384(7)   \\
6.240   & 1.02     &  0.448(6)    & 0.63(5)      & 0.417(11)  \\
6.218   & 1.06     &  0.444(9)    & 0.56(5)      & 0.448(21)  \\
6.136   & 1.19     &  0.43(7)     & 0.46(6)      & 0.51(4)    \\
\bottomrule
\end{tabularx}
\end{center}
\end{table*}

\begin{figure}[htb] 
\centering
\includegraphics[width=1.0\linewidth,clip]{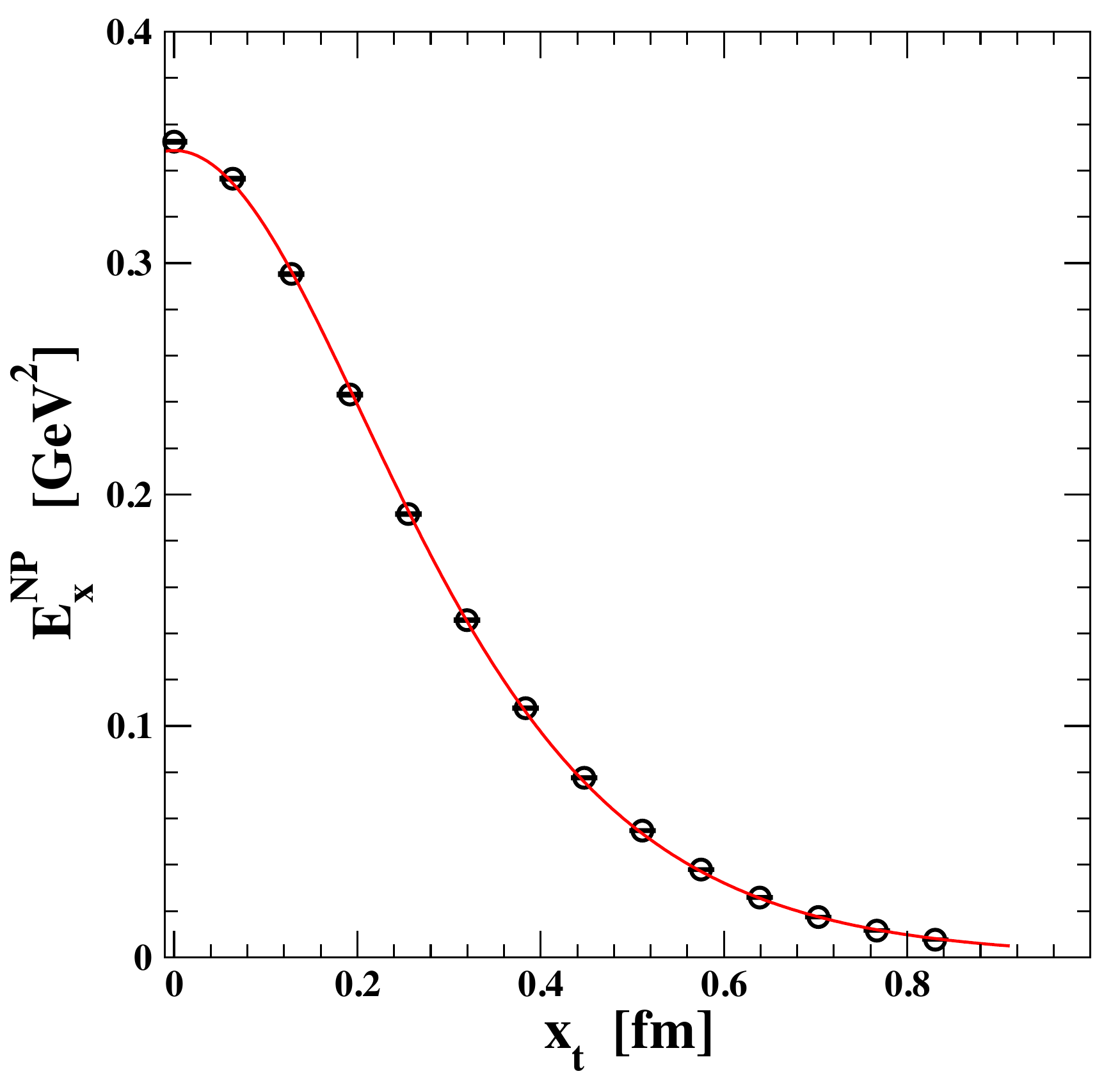}
\caption{The Clem fit (Eq.~(\ref{clem2}) to the non-perturbative chromoelectric field $E_x^{\rm NP}$ obtained from the curl procedure
for $\beta=6.240$, $d=0.51 \, {\mathrm {fm}}$, $x_l=3a$.
}
\label{fig:clemsamplefit}
\end{figure}
\begin{figure}[htb] 
\centering
\includegraphics[width=1.0\linewidth,clip]{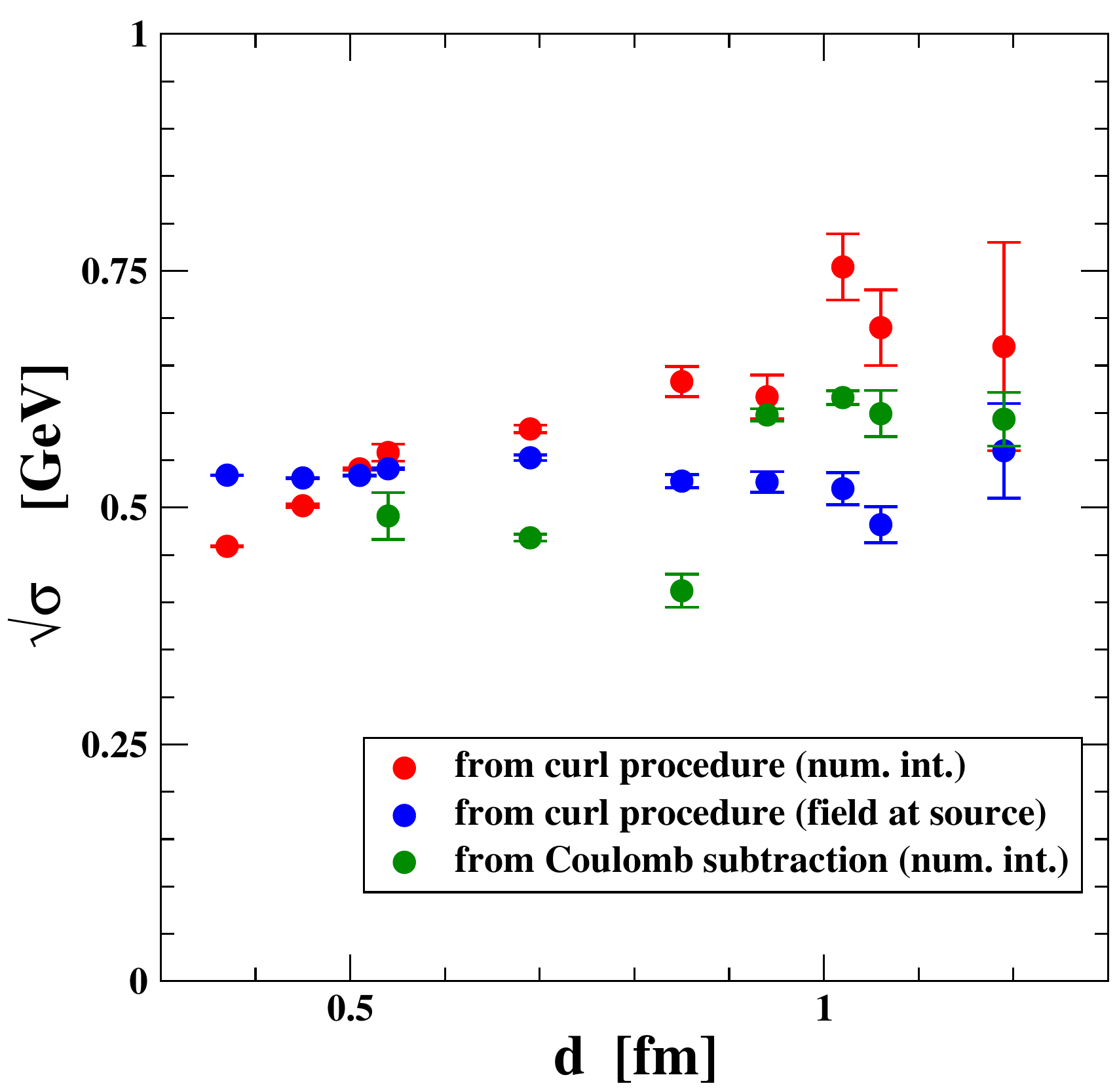}
\caption{The square root of the string tension obtained using several different procedures:
(i) by numerical integration of Eq.~((\ref{sqrtstring-clem}) with the non-perturbative field obtained from the
curl procedure (Section~\ref{sect:nonpert}); 
(ii) from the non-perturbative field obtained from the curl procedure evaluated and evaluated at sources;
(iii) by numerical integration of Eq.~(\ref{sqrtstring-clem}) with the non-perturbative field obtained from the
Coulomb subtraction (Eq.~\eqref{ENP}).
}
\label{fig:sqrtstring}
\end{figure}
\begin{figure}[htb] 
\centering
\includegraphics[width=1.0\linewidth,clip]{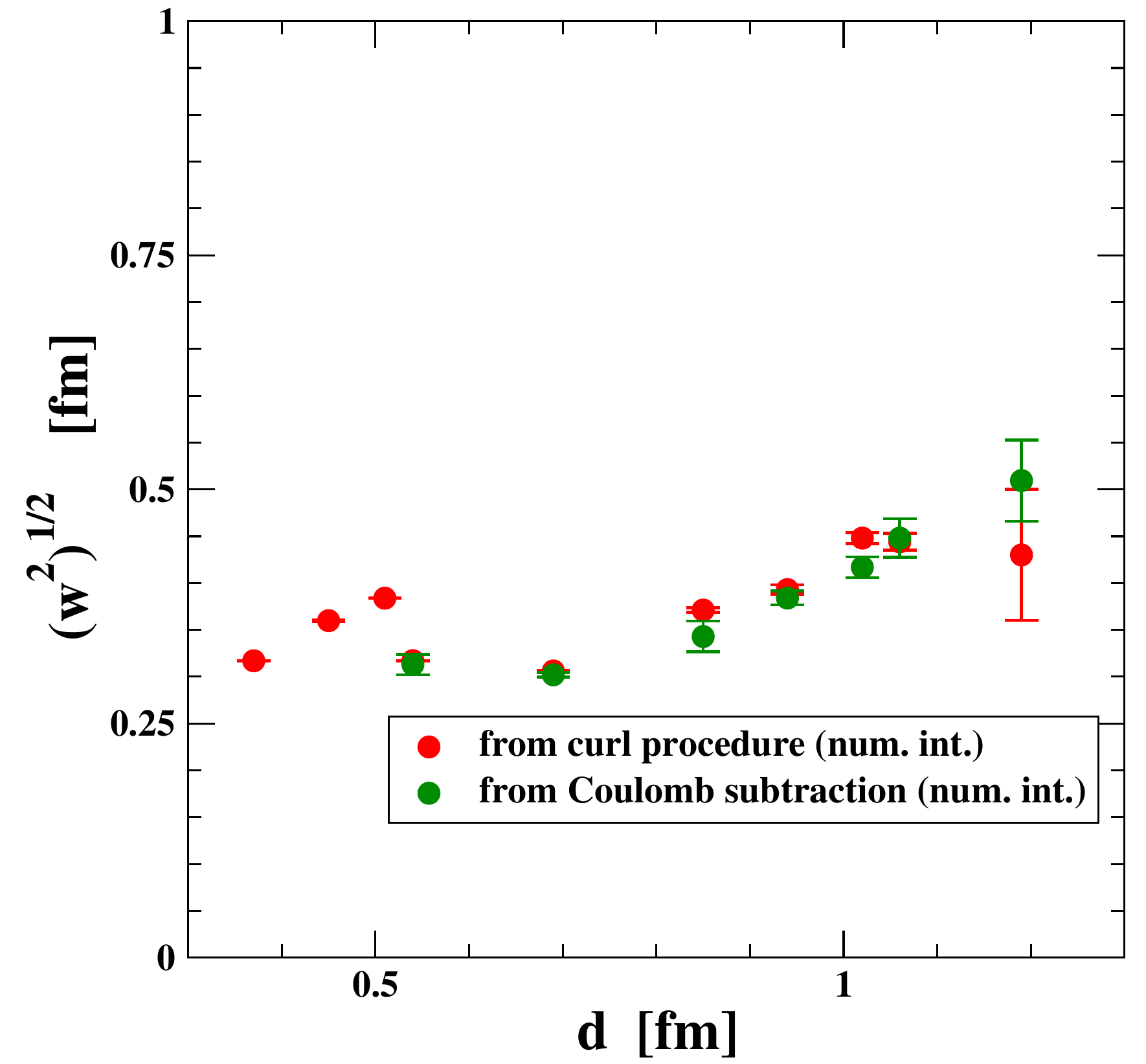}
\caption{The mean square width of the flux tube obtained using several different procedures:
(i) by numerical integration of Eq.~(\ref{width}) with the non-perturbative field obtained from the
curl procedure (Section~\ref{sect:nonpert}); 
(ii) by numerical integration of Eq.~(\ref{sqrtstring-clem}) with the non-perturbative field obtained from the
Coulomb subtraction (Eq.~\eqref{ENP})
}
\label{fig:width}
\end{figure}

\section{The string tension and the width of the flux tube}\label{sec:strTensAndWidth}

The forces between  charged particles in  electrodynamics 
are determined by a stress tensor $T_{\mu \nu}$
constructed from  fields  $F_{\mu \nu}$ satisfying Maxwell's equations (see Eq.  (12.113) in Ref.~\cite{jackson_classical_1999}).
 Similarly, the force
between quarks and antiquarks in Yang Mills theory
is determined by
the  stress tensor $T_{\mu \nu}$,  Eq. (\ref{Tmunu}), constructed
from the field tensor $F_{\mu \nu}$  obtained from our simulations.

The quark-antiquark force $\vec{F}$ is  then the integral of the longitudinal component 
$T_{x x} = (E_x (x_l = d/2,  x_t))^2/2$ of the stress tensor over the median plane $x = d/2$
bisecting the line connecting the quarks, Eq.~\eqref{newforce}. The non-perturbative quark-antiquark force $\vec{F}_{\rm NP} = -\hat{e}_x \sigma$ determining the string tension $\sigma$  has the corresponding expression in terms of the non-perturbative longitudinal component of 
the stress tensor
$T^{\rm NP}_{x x} = (E^{\rm NP}_x (x_l = d/2,  x_t))^2/2 \equiv (E_x^{\rm NP})^2(x_t))/2.$  

The square root of the string tension is then equal to
\begin{equation}
\label{sqrtstring-clem}
\sqrt{\sigma} =  \sqrt{ \int d^2x_t 
  \frac{(E^{\rm NP}_x)^2 (x_t)}{2} } .
\end{equation}
We have evaluated the integral (Eq.~(\ref{sqrtstring-clem}) ) in two ways:
\begin {enumerate}
\item  by direct numerical integration, using the values of $E_x^{\rm NP}$ 
determined by our simulations, and
\item  analytically,
by fitting the numerical data 
for the transverse distribution of $E_x^{\rm NP} (x_t)$  as in~\cite{Cardaci:2010tb,Cea:2012qw,Cea:2013oba,Cea:2014uja,Cea:2014hma} to the Clem 
parameterization of the field surrounding a magnetic vortex in a superconductor \cite{Clem:1975aa}.
\begin{equation}
\label{clem2}
E_x^{\rm NP} (x_t)  =  \frac{\phi}{2 \pi} \frac{\mu^2}{\alpha} \frac{K_0[(\mu^2 x_t^2 
+ \alpha^2)^{1/2}]}{K_1[\alpha]} \; ,
\end{equation}
where $\phi, \mu$ and $\alpha$ are fitting parameters. In the dual superconducting model~\cite{'tHooft:1976ep,Mandelstam:1974pi,Ripka:2003vv,Kondo:2014sta}
$\lambda = \frac{1}{\mu}$ is the penetration depth and
 
\begin{equation}
\label{landaukappa}
\kappa =  \frac{\sqrt{2}}{\alpha} 
\left[ 1 - K_0^2(\alpha) / K_1^2(\alpha) \right]^{1/2} \,,
\end{equation}
is the Landau-Ginzburg parameter characterizing the type of superconductor.
\end{enumerate}

Figure (\ref{fig:clemsamplefit}) shows an example of the fit of the data  to the Clem functional form Eq.~(\ref{clem2}).
for the transverse distribution of  
$E_x^{\rm NP} (x_t)$,
obtained using the {\em{curl}} procedure.

We  can obtain a second expression for the string tension by utilizing the result ~\cite{Brambilla:2000gk,brambilla2019static} 
that the force on a quark is equal to the value of the chromoelectric field
 at the position of the quark. The string tension is then equal to the 
corresponding value of the confining part of the chromoelectric field
\begin{equation}
\sigma = E^{\rm NP}_x (x_l =0, x_t =0).
\label{sigma}
\end{equation}
Eqs.~\eqref{sqrtstring-clem} and \eqref{sigma} provide two independent ways to extract the string tension from 
simulations. As mentioned earlier, we must use the curl method to isolate the confining field 
to extract the string tension by Eq.~(\ref{sigma}).

 To obtain additional information
 about the structure of the chromoelectric flux tube we have calculated
the {\it  mean square  root width}:
 \begin{equation}
\label{width}
\sqrt{w^2} = \sqrt{\frac{\int d^2x_t \, x_t^2 E_x(x_t)}{\int d^2x_t \, E_x(x_t)}}.
\end{equation}
Just as we have evaluated  the integral (\ref{sqrtstring-clem}) for the string tension, we have
evaluated  the integral (\ref{width}) for the mean square root width both
numerically, using the data for $E_x^{\rm NP} (x_t)$, and analytically, in terms
of Clem parameters, fitting the longitudinal component of $E_x^{\rm NP}(X_t)$
 in the median plane to the Clem parametrization
 (\eqref{clem2}). 
The results of that fit are given in
Table~\ref{tab:clem_parameters}.
In most cases the parametrization in Eq.~\eqref{clem2} gives a good description of
the field, shown by the  values of $\chi^2_r$, though
the parameters themselves are somewhat unstable, which reflects the strong
correlation between the parameter estimates.

We compared two different methods for calculating the
 integrals in Eqs.~ (\ref{sqrtstring-clem}) and (\ref{width}).
First, we carried out the numeric integration
in Eqs.~(\ref {sqrtstring-clem}) and (\ref{width}), respectively, postulating the
rotational symmetry of the field. This approach was repeated for 
both the non-perturbative field obtained using the ``curl procedure''
(resulting in $\sqrt{\sigma_{\rm int}}$ and $\sqrt{w^2_{\rm int}}$) 
and the field obtained using Coulomb subtraction (resulting in 
$\sqrt{\sigma_{\rm Coulomb}}$ and $\sqrt{w^2_{\rm Coulomb}}$). Next we calculated 
the values of the string tension and the mean square root width using
the Clem parameters given in Table~\ref{tab:clem_parameters} to get the values
denoted as $\sqrt{\sigma_{\rm Clem}}$ and $\sqrt{w^2_{\rm Clem}}$. 

(One remark should be made for the width -- while we know that the value of 
$E_x^{\rm C} (x, y_{\mathrm{max}} + 1)$ in Eq.~\eqref{curl_Ex} is small
($O(y_{\mathrm{max}}^{-2})$), in the numerator of Eq.~\ref{width} this small
constant will be multiplied by $y^2$ ($y^3$ after the integration over polar
angle), which will cause the error introduced to increase with $y_{\max}$. 
Indeed, the comparison with the analysis done taking $E_x^{\rm NP}(x, y_{\rm max})
= 0$ shows large discrepancies in this case.)

Finally, we evaluated  expression (\ref{sigma}) for the
string tension, $\sigma = E^{\rm NP}_x (x_l =0, x_t =0)$,  using the curl method to
determine the  magnitude of the non-perturbative field at the
sources.

 Our results   are gathered in Tables~\ref{tab:string_tension}
and~\ref{tab:width}, where we use the notation $\sigma_0 \equiv  E^{\rm NP}_y (x_l =0, ~x_t =0)$
and in Fig.~(\ref{fig:sqrtstring}). 
The data shown in  Fig.~(\ref{fig:sqrtstring}) give a consistent value
of $\sqrt{\sigma}$ for all values of the separation $d$, with  scatter that increases
with $d$ as the resolution diminishes.
The values of $\sqrt{\sigma}$ lie close to $ 0.465$ GeV, the value used in the parameterization
 Ref.~\cite{Necco:2001xg}

Let us review the basis of our calculations. 
Our hypothesis is 
that the string tension  is determined by the 
field $\vec{E}$ we measured (the 'Maxwell' mechanism). We have determined $\sigma$  from both the transverse 
structure of the flux tube (Eq. ~\eqref {sqrtstring-clem})
and  its longitudinal structure (Eq.~\ref{sigma}) )
as shown in Fig.~(\ref{fig:Field_nonperturbative_Ex}), in which
the non-perturbative field has been isolated.

We emphasize that, as discussed in Section (\ref{sec:distrColFields}), 
the 'Maxwell' mechanism cannot be used
to obtain the Coulomb correction to the string tension.
 This implies that the fluctuating
fields not measured  in our simulations must contribute to the Coulomb force.

On the other hand, the Coulomb correction has been obtained by recent direct
simulations of the stress energy-momentum tensor 
in Yang Mills theory
\cite{Yanagihara:2018qqg}.  The Yang Mills stress tensor 
accounts for the contributions of fluctuating
fields but cannot be directly related to measured fields,   
in contrast to the Maxwell stress tensor
$T_{xx} = (1/2) E_x^2$,  determining the string tension. 
 (See Eq.~(\ref{components}).) 
 
\section{Conclusions and outlook}\label{sec:conclAndOutlook}

In this paper we have determined the spatial distribution in three dimensions of
all components of the color fields generated by a static quark-antiquark pair. 
We have found that the dominant component of the color field is the
chromoelectric one in the longitudinal direction, {\it i.e.} in the
direction along the axis connecting the two quark sources. This feature
of the field distribution has been known for a long time. However, the
accuracy of our numerical results allowed us to go far
beyond this observation. First, we could confirm that, as observed
in~\cite{Baker:2018mhw}, all the chromomagnetic
components of the color field are compatible with zero within the statistical
uncertainties. Second, the chromoelectric components of the color fields
in the directions transverse to the axis connecting the two sources,
though strongly suppressed with respect to the longitudinal component,
are sufficiently greater than the statistical uncertainties that they
can be nicely reproduced  by a Coulomb-like field generated
by two  sources with opposite charge (everywhere except in a small region around
the sources).

In Ref.~\cite{Baker:2018mhw} we subtracted this Coulomb-like field from the
simulated chromoelectric field to obtain a non-per\-tur\-ba\-tive
field $\vec{E}^{\rm NP}$
according to Eq.~\eqref{ENP} and found that the dependence of the resulting
longitudinal component of  $\vec{E}^{\rm NP}$ on the distance $x_t$ from the axis
is independent of the position $x_l$ along the axis, except near the sources,
thus suggesting that the non-perturbative field found in this way from lattice
simulations can be identified as the confining field of the QCD flux tube.

In this work we have improved the approach of Ref.~\cite{Baker:2018mhw} by
presenting a new procedure to subtract the Coulomb-like field, which does
not rely on any preconception about its analytic form, but is based only
on the requirement that its curl is equal to zero.

Moreover, we have carefully analyzed the spatial distribution of the subtracted,
non-perturbative part of the longitudinal chromoelectric field to extract from
it some relevant parameters of the flux tube, such as the mean width and the
string tension, both by means of a fully numerical, model-independent procedure
and by a prior interpolation with the dual version of the Clem function for
the magnetic field in a superconductor.

We have also used our determinations of the color field components to
construct the `Maxwell' stress tensor. Details about its determination
and a comparison with the recent literature about this
topic~\cite{Yanagihara:2018qqg} are presented in~\ref{AppendixStressTensor}.

In conclusion, we have shown that the separation of the chromoelectric field
into perturbative and non-perturbative components can be obtained by directly
analyzing lattice data on color field distributions between static quark
sources, with no need of model assumptions. To the best of our knowledge,
this separation between perturbative and non-per\-tur\-bative components has not
been carried out previously. It  provides  new understanding of the
chromoelectric field surrounding the quarks. We have used the non-perturbative field to calculate 
the string tension and the spatial distribution of the energy-momentum tensor surrounding the 
static quarks, under the assumption that the fluctuating color
 fields not measured in our simulations do not contribute to the string tension.  
 The extension of our approach
to the case of QCD with dynamical fer\-mions with physical masses and at
non-zero temperature and baryon density is straightforward~\cite{Cea:2015wjd}.

\appendix

\section{The `Maxwell'' stress tensor}\label{AppendixStressTensor}

In this Appendix we consider the ``Maxwell'' energy-mo\-men\-tum tensor
$T_{\mu \nu}$ as a function of the field tensor $F_{\mu \nu}$ characterizing the
SU(3) flux tube, which is in its turn defined in terms of the gauge invariant
correlation function $\rho^{\rm conn}_{W, \mu\nu}$ of Eqs.~(\ref{rhoW})
and~(\ref{rhoWlimcont}) and points in a  single color direction parallel to
the color direction of the source (which is determined dynamically).
Its six tensor components (the electric and
magnetic fields $\vec{E}$ and $\vec{B}$) correspond to the six orientations of
the plaquette $U_P$ relative to that of the Wilson loop
(see Fig.~\ref{fig:operator_Wilson}).

The simulated fields $\vec{E}$ and $\vec{B}$ have the space-time symmetries of
the Maxwell fields of electrodynamics, while carrying color charge in a
single direction in color space. 
The energy-momentum tensor $T_{\mu \nu}$ lies in the same direction in color
space of the simulated fields $\vec E$ and $\vec B$ and has the (Euclidean)
Maxwell form:
\begin{equation}
T_{\mu\nu} = F_{\mu \alpha} F_{\alpha \nu} - g_{\mu \nu} F_{\alpha \beta}F_{\alpha \beta}/4 \;.
\label{Tmunu}
\end{equation}
Its spatial components $\mu=i, \nu=j$, with $i,j=1,2,3$ determine the Maxwell
stress tensor:
\begin {equation}
T^{\rm Maxwell}_{ij} = -T_{ij}\;. 
\label {TMaxwell} 
\end{equation}
Taking $\mu =i$ and $\nu = j\neq i$ in Eq.~\eqref{Tmunu} gives
\begin {equation}
T_{ij}^{\rm Maxwell} =  -T_{ij} = E_{i}E_{j} + B_iB_{j} - \delta_{ij} (E^2 + B^2)/2\;,
\label{Tij} 
\end{equation}
while the diagonal time component $-T_{44}$ of $T_{\mu \nu}$ determines the
energy density,
\begin {equation}
-T_{44} = \frac{1}{2} (E^2 + B^2) \;.
\label{T44}
\end{equation}
         
We use cylindrical coordinates $(x,~r,~\theta)$, $r\equiv\sqrt{y^2 + z^2}$,
$\tan \theta \equiv z/y$, and the corresponding unit vectors $\hat{e}_r$, 
$\hat{e}_\theta$: 
\begin {eqnarray}
\label{er}
\hat{e}_r &=& \hat{e}_y \cos \theta + \hat{e}_z \sin \theta\;, \\
\label{etheta}
\hat{e}_{\theta} &=& -\hat{e}_y \sin \theta + \hat{e_z} \cos \theta 
\end{eqnarray}
($x$ is the longitudinal direction of the flux tube, {\it i.e.} the axis
along which the static sources are located).

The force exerted by the antiquark on the quark can be expressed, by means of
the stress tensor, as a surface force $\vec{F}$ acting on the infinite plane
$x = \frac{d}{2}$ bisecting the line connecting the pair:
\begin {equation} 
\vec{F}_j =  \int \int dy~dz~\hat n_i  T^{\rm Maxwell}_{ij} (x=d/2,~y, ~z) \;,
\label{force}
\end{equation}
where $\hat{n} = - \hat{e}_x$ is the outward  normal to the region $x > \frac{d}{2}$
containing the quark.  The only non-vanishing component of  the quark-antiquark force $\vec{F}$  is longitudinal, so
\begin {equation} 
\vec{F} = - \int \int dy~dz~\hat {e}_x  T_{xx} (x=d/2,~y, ~z).
\label{newforce}
\end{equation}
Using the components in Eq.~\eqref{Tij} of $T_{ij}^{\rm Maxwell}$, 
and taking into account that the measured magnetic field $B$ is compatible with zero
\begin {eqnarray}
\label{components}
-T_{xx} &=& \frac{1}{2} (E_x^2 - E_r^2),\;\;\;\;\; E_r^2=E_y^2 + E_z^2 \;, 
\nonumber \\
-T_{xy} &= &E_x E_y \;, \\
-T_{xz} &=& E_x E_z \;, \nonumber
\end{eqnarray}
in Eq.~\eqref{force} gives
\begin {equation}
  \vec{F} = - \int_o^{2 \pi} d \theta \int_0^\infty r ~dr
      [ \hat{e}_x \frac{(E_x^2 - E_r^2)}{2} + \hat{e}_r E_x E_r]\;.
\label{forceagain}
\end{equation}
The angular average over the radial vector $\hat{e}_r$ in Eq.~\eqref{forceagain}
vanishes. Furthermore by symmetry the transverse field $E_r$ on the mid-plane
$x=\frac{d}{2}$ vanishes, so that the quark-antiquark force in Eq.~\eqref{forceagain} becomes
\begin{equation}
\vec{F} = - 2 \pi \int_0^\infty r dr \frac {E_x^2 (r)}{2} \hat{e}_x\;.
\label{finally}
\end{equation}
Replacing $E_x (r)$ by the non-perturbative field $E_x^{\rm NP}(r)$
in Eq.~\eqref{finally} gives the non-perturbative quark-antiquark force
$\vec{F}_{\rm NP}$,
\begin {equation}
  \vec{F}_{\rm NP} = - \sigma \hat{e}_x\;, \;\;\;
  \sigma = 2 \pi \int_0^\infty r dr \frac{(E_x^{\rm NP} (r))^2}{2}\;.
\label{stringtension}
\end{equation}
Eq.~(\ref{stringtension}) determines the string tension $\sigma$ in terms of
the longitudinal component of the non-perturbative field $E_x^{\rm NP} (r)$, the
confining component of the SU(3) flux tube. We have already presented it,
in a slightly different notation, in Eq.~\eqref{sqrtstring-clem}.

Using Eqs.~\eqref{er} and~\eqref{etheta} in Eq.~\eqref{Tij} we can obtain the
components of the Maxwell stress tensor in cylindrical coordinates:
\begin{eqnarray}
\label{Trr}
-T_{rr} (x,r)&=& -(\hat{e}_r)_i T_{ij} (\hat{e_r})_j~=~-\frac{1}{2}
(E_x^2 - E_r^2)\;,\\
\label{Tthetatheta}
-T_{\theta \theta} (x,r)&=& -(\hat{e}_\theta)_i T_{ij}
(\hat{e_\theta})_j~=~-\frac{1}{2} (E_x^2 + E_r^2)\;,\\
\label{Txx}
-T_{xx} (x,r)&=& -(\hat{e}_x)_i T_{ij} (\hat{e_r})_j~=~~~~~\frac{1}{2}
(E_x^2 - E_r^2)\;,\\
\label{Txr}
-T_{xr} (x,r)&=& -(\hat{e}_x)_i T_{ij} (\hat{e_r})_j~=~~~~E_xE_r\;.
\end{eqnarray}
The remaining non-vanishing component of $T_{\mu\nu}$ is the energy density
$T_{44}$,
\begin{eqnarray}
\label{T44new}
-T_{44}(x,r)&=& \frac{1}{2} (E_x^2 + E_r^2)\;.
\end{eqnarray}

Eqs.~(\ref{Trr})-(\ref{T44new}) express all components of the stress tensor
in terms of the simulated color fields $E_x (x, r)$ and $E_r(x, r)
=\sqrt{E_y^2 + E_z^2}$. On the symmetry plane $x=\frac{d}{2}$, $E_r=0$ and
Eqs.~(\ref{Trr})-(\ref{T44new}) reduce to
\begin{eqnarray}
\label{T44short}
T_{44} (r) &=& T_{xx}(r) = -\frac{E_x^2(r)}{2}\;,\\
\label{Trrshort}
T_{rr} (r) &= &T_{\theta\theta} (r)  ~= \frac{E_x^2(r)}{2}\;.
\end{eqnarray}
Further, we note that the trace of the stress tensor $T_{\mu \nu}$ evaluated
from Eqs.~\eqref{Trr}-\eqref{T44new} vanishes independently of the simulated
flux-tube fields $E_x (x, r)$ and $E_r (x, r)$:
\begin {equation}
T_{44} (x,r) + T_{xx} (x, r) + T_{rr} (x,r) + T_{\theta\theta} (x, r) =0\;.
\label{trace}
\end{equation}

\begin{figure}[H]
\centering
\includegraphics[width=0.8\linewidth,clip]{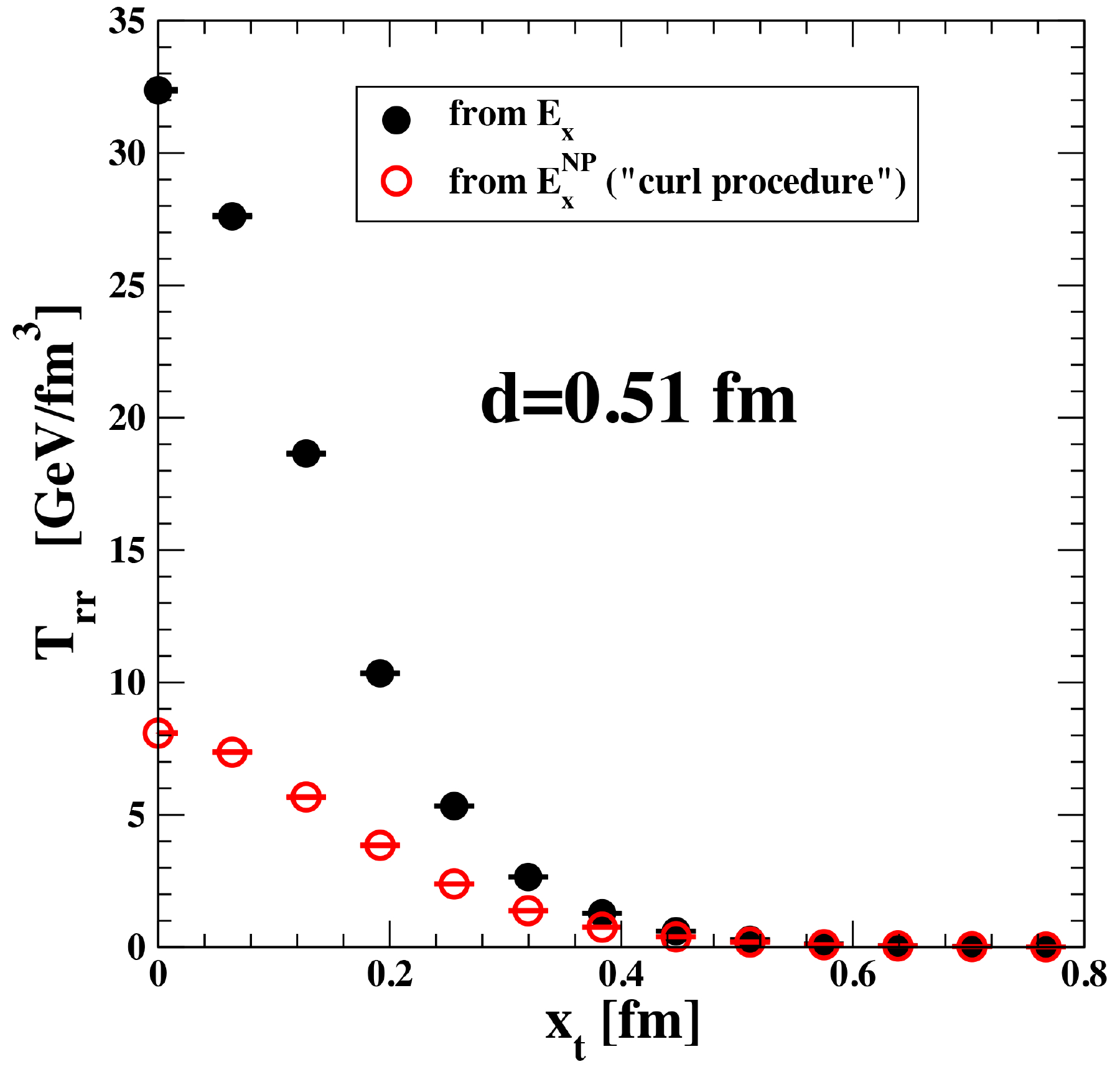} \\
\includegraphics[width=0.8\linewidth,clip]{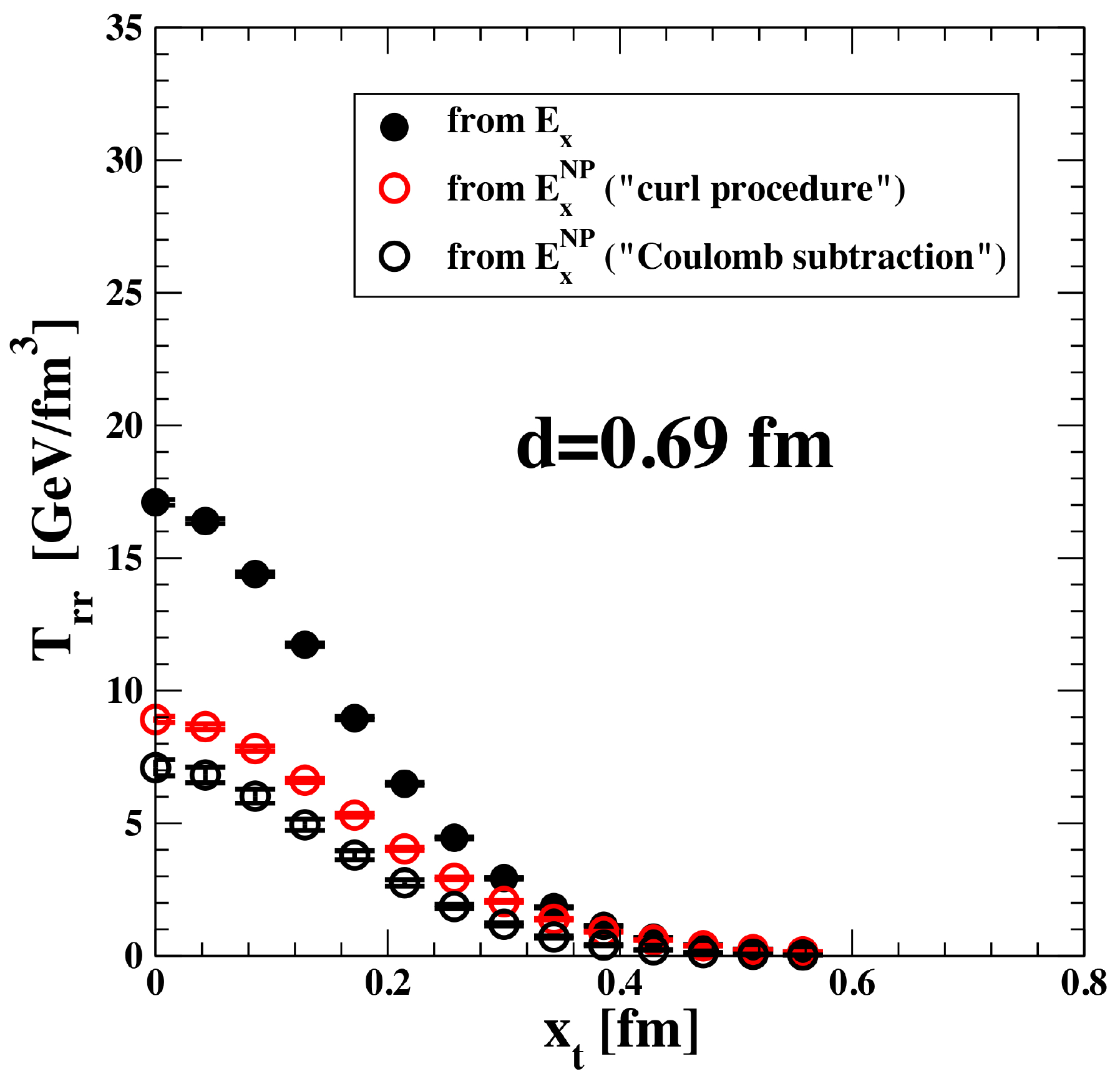} \\
\includegraphics[width=0.8\linewidth,clip]{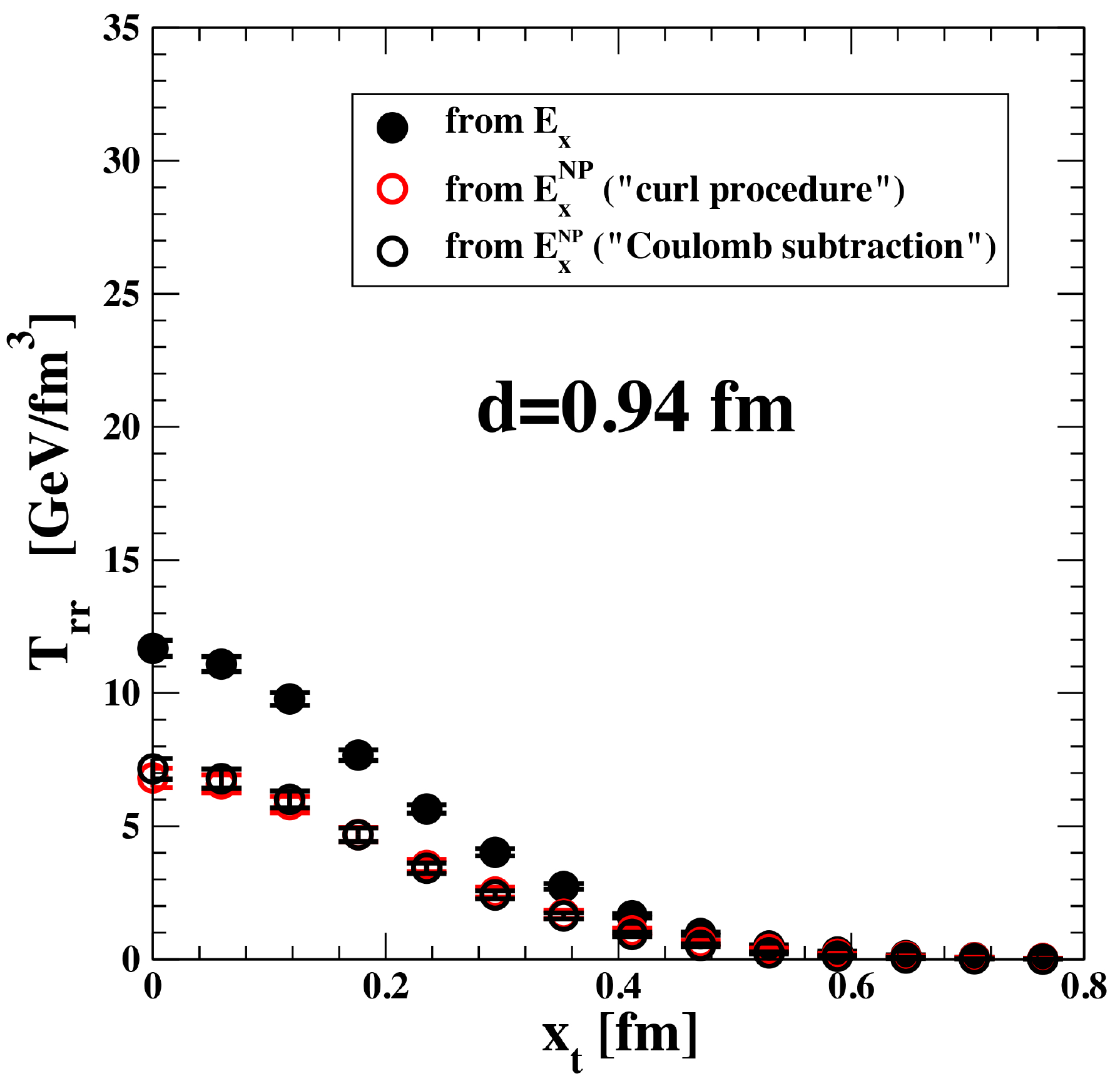}
\caption{The diagonal components of the Maxwell stress tensor recovered from the 
full field $E_x$ (filled circles) and non-perturbative field $E_x^{(\rm NP)}$ (empty circles) 
for $d=0.51$ fm (top), $d=0.69$ fm (middle), and $d=0.95$ fm (bottom).
}
\label{fig:Maxwell_stress_tensor}
\end{figure}

We have calculated the non-perturbative content of $T_{rr}$ on the symmetry
plane (where $T_{rr}=T_{\theta\theta}=-T_{44}=-T_{xx}$) versus $r$ for three
different values of the quark-antiquark distance:
$d=0.51$~fm (at $\beta=6.240$), $d=0.69$~fm (at $\beta=6.539$)
and $d=0.95$~fm (at $\beta=6.299$). Results are presented in
Fig.~\ref{fig:Maxwell_stress_tensor}, where also the
full (non-perturbative plus Coulomb) content of $T_{rr}$ is shown.

  The width of the energy density distribution $T^{\rm NP}_{44}$ can be
  obtained through Eq.~(\ref{width}), with $E^{\rm NP}_x$ replaced by $T^{\rm NP}_{44}$ as
  given in Eq.~\eqref{T44short}; results are presented in
  Table~\ref{tab:tensor_width}.  Since $T_{44}^{\rm NP}$ 
  is proportional to $\left(E_x^{\rm NP}\right)^2$ the width of the 
  $T_{44}^{\rm NP}$ component of the Maxwell stress tensor obtained
   from the nonperturbative field 
  given in Table~\ref{tab:tensor_width} is systematically smaller 
  than the width of the nonperturbative part of the longitudinal chromoelectric 
  field component $E_x^{\rm NP}$ 
  given in Table~\ref{tab:width}.  (The square of the field  decreases more rapidly 
with distance than the field itself.)

We now compare the above results  obtained using our measured
flux tube fields to evaluate the 'Maxwell' energy-momentum tensor 
$T_{\mu \nu}$ with the corresponding results of recent direct simulations
\cite{Yanagihara:2018qqg} of the expectation value of
the energy momentum tensor $T_{\mu\nu}^{\rm YM}$
in the presence of a quark-antiquark pair. The latter simulations, which measure the energy
and stresses in all color directions directly, were carried out in the plane midway between the quark and the antiquark, for three values of their separation.

The tensor  $T_{\mu\nu}^{\rm YM}$ has the form \cite{Suzuki:2013gza}
\begin{equation}
  T_{\mu \nu} = F_{\mu \alpha}^a F_{\alpha \nu}^a - g_{\mu \nu} F_{\alpha \beta}^a
  F_{\alpha \beta}^a/4\,,
   \label{Tmunua}
\end{equation}
where $F_{\mu \nu}^a$ is the Yang-Mills field tensor in the adjoint
representation of SU(3),
\begin{equation}
\label{Fmunua}
F_{\mu \nu}^a = \partial_\mu A_\nu^a - \partial_\nu A_\mu^a + g f_{abc} A_\mu^b
  A_\nu^c\,,
\end{equation}
where $f_{abc}$ are the structure constants of the SU(3) algebra. In our
definition the field is squared after color projection, whereas
in Eq.~\eqref{Tmunua} the sum over color components is taken after 
squaring. Moreover, the stress tensor in
Ref.~\cite{Yanagihara:2018qqg} is renormalized (this motivates the 
superscript $R$
in the formulas below).

%
\begin{figure}[htb]
\centering
\includegraphics[width=0.8\linewidth,clip]{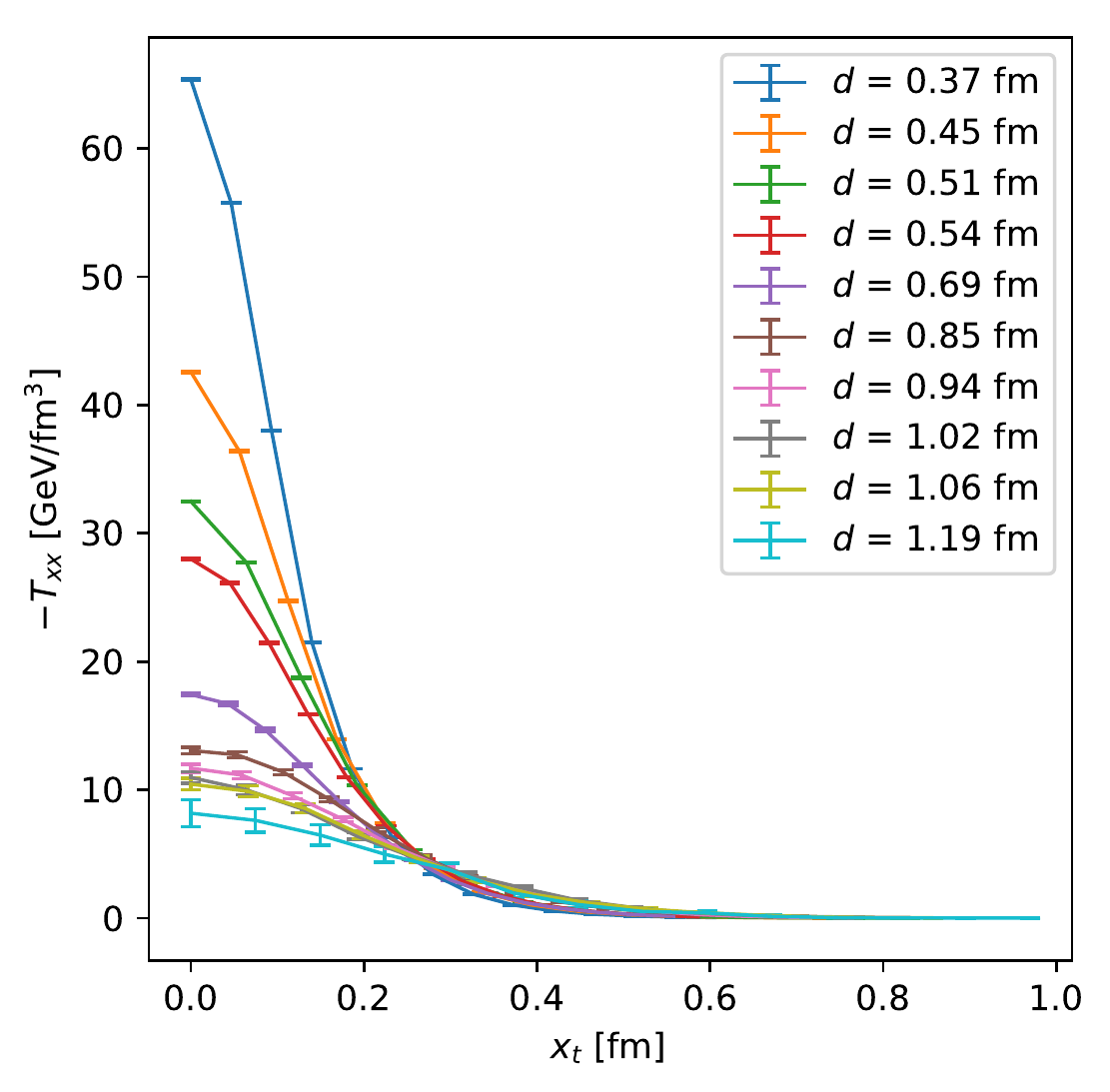} \\
\includegraphics[width=0.8\linewidth,clip]{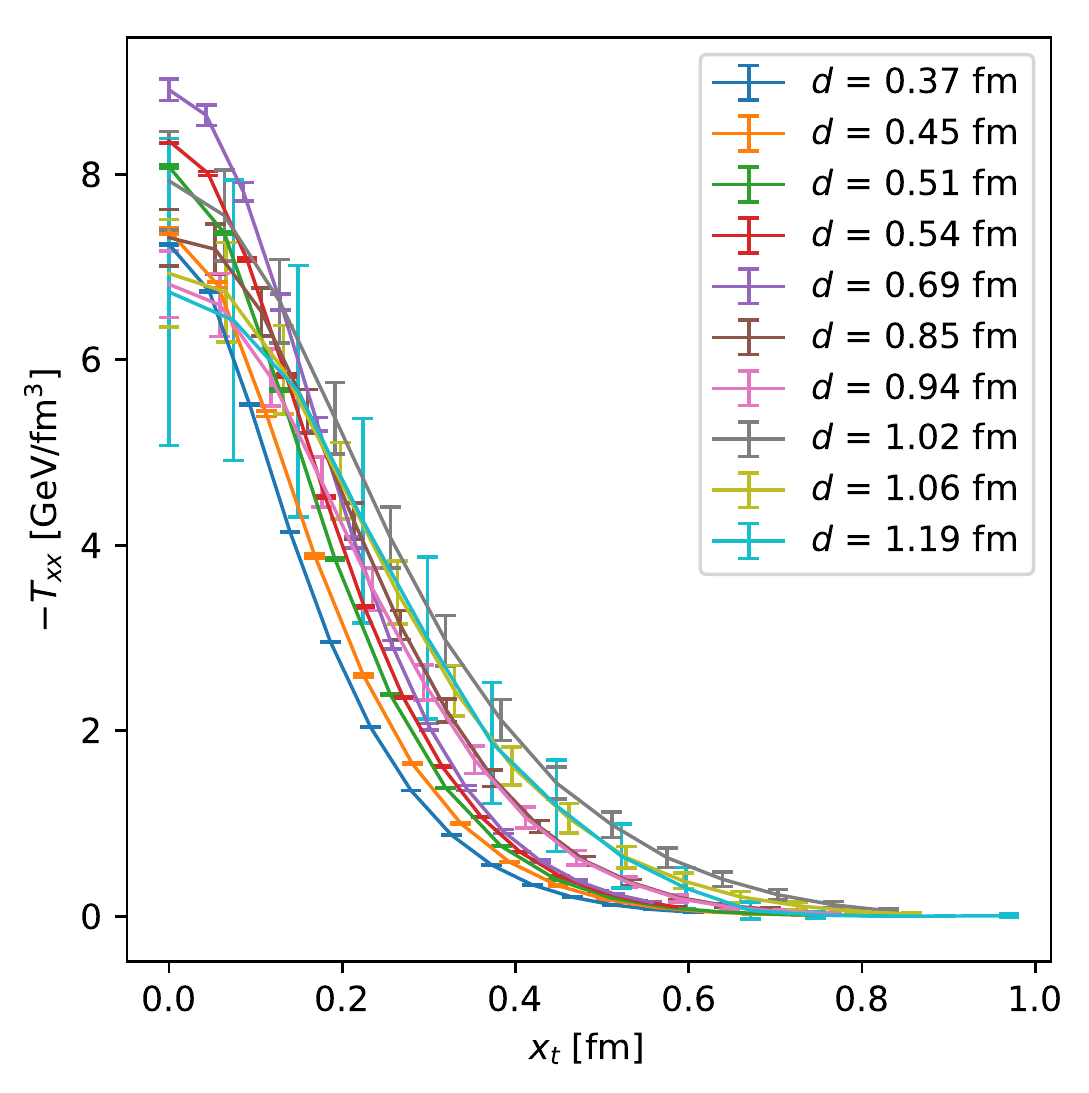} 
\caption{Comparison of the diagonal components of the ``Maxwell'' stress tensor 
for the different quark-antiquark separations, recovered from the 
full field $E_x$ (top) and from the non-perturbative field $E_x^{(\rm NP)}$ (bottom).
}
\label{fig:Maxwell_stress_tensor_comparison}
\end{figure}
\begin{table}[htb]
\begin{center} 
  \caption{The width of the diagonal component of the Maxwell stress tensor 
  recovered from the full field $E_x$ ($\sqrt{w^2_{\rm Full}}$) 
  and non-perturbative field $E_x^{(\rm NP)}$ ($\sqrt{w^2_{\rm NP}}$).}
  \label{tab:tensor_width}
\begin{tabularx}{0.382\textwidth}{@{} |S[table-format=1.5]|S[table-format=1.2]|S[table-format=1.9]|S[table-format=1.9]| @{}}
\toprule
$\beta$ & {$d$ [fm]} & {$\sqrt{w^2_{\rm Full}}$ [fm]} & {$\sqrt{w^2_{\rm NP}}$ [fm]} \\ 
\midrule
 6.47466 & 0.37 & 0.18751(4)  & 0.24438(9)  \\
 6.333   & 0.45 & 0.2148(3)   & 0.2637(7)   \\
 6.240   & 0.51 & 0.23014(10) & 0.26925(27) \\
 6.500   & 0.54 & 0.22597(15) & 0.25572(21) \\
 6.539   & 0.69 & 0.2360(6)   & 0.2511(8)   \\
 6.370   & 0.85 & 0.2833(25)  & 0.298(3)    \\
 6.299   & 0.94 & 0.300(3)    & 0.306(6)    \\
 6.240   & 1.02 & 0.326(6)    & 0.360(9)    \\
 6.218   & 1.06 & 0.326(7)    & 0.345(10)   \\
 6.136   & 1.19 & 0.320(20)   & 0.319(27)   \\
\bottomrule
\end{tabularx}
\end{center}
\end{table}

In \cite{Yanagihara:2018qqg}  the expectation value of the energy-momentum tensor in the background of a quark-antiquark pair  is denoted by  $ \left\langle ~T_{ij}^R (r) ~\right\rangle_{Q \bar{Q}} $. We will use this notation in comparing our results (\ref{T44short}) and (\ref{Trrshort}) with that work. 

 In  \cite{Yanagihara:2018qqg}   the $r$ dependence of the components of $\left\langle~T_{\mu\nu}^R(r)~\right\rangle_{Q\bar{Q}}$ was plotted for the 3 values of the quark-antiquark separation for which simulations were made, and these 'noticeable' features of the results were pointed out:

\begin{enumerate}
\item Approximate degeneracies between temporal and longitudinal components  and between radial and angular components are found for a wide range of $r$;
\begin{eqnarray}
&\left\langle T_{44}^R(r)\right\rangle_{Q\bar{Q}} ~\approx ~\left\langle T_{xx }^R(r)\right\rangle_{Q\bar{Q}} ~ > ~0\,,\nonumber\\
&\left\langle T_{rr}^R(r)\right\rangle_{Q\bar{Q}}~ \approx ~ \left\langle T_{\theta\theta}^R(r)\right\rangle_{Q\bar{Q}}~  >~ 0.
\label{T44rrdegen}
\end{eqnarray}
We emphasize that the two inequalities in Eq. (\ref{T44rrdegen}) are  general consequences of (\ref{T44short}) and (\ref{Trrshort}), independent of the values of the simulated field $E_x (x,r)$. In
contrast, a recent  study \cite {Yanagihara:2019foh} of the stress  tensor  distribution in the Abelian
Higgs model found that these relations could only be satisfied within a very
narrow range of the model parameters .

\item The scale symmetry broken in the YM vacuum (the trace anomaly),
\begin{eqnarray}
&\left\langle T_{44}^R(r)\right\rangle_{Q\bar{Q}} + \left\langle T_{xx}^R(r)\right\rangle_{Q\bar{Q}}&\\ &+ \left\langle T_{rr}^R(r)\right\rangle_{Q\bar{Q}} + \left\langle T_{\theta\theta}^R(r)\right\rangle_{Q\bar{Q}}&~ < ~0,\nonumber
\label{tracedegen}
\end{eqnarray}
is partially restored inside the flux tube.

\item Each component of the energy-momentum tensor at r=0 decreases as the separation
 becomes larger, while the transverse radius of the flux tube, 
 typically about 0.2 fm, seems to increase for large separations
\cite{Luscher:1980iy, Gliozzi:2010zv, Cardoso:2013lla}, although
 the statistics are not sufficient 
to discuss the radius quantitatively. 

 We see some indication
 of the increase in the width
of the distributions of the diagonal components of the 'Maxwell' stress tensor in Fig.
(\ref{fig:Maxwell_stress_tensor_comparison}) and Table (\ref{tab:tensor_width})
for all ten values of the quark-antiquark separation.
However, this width is greater than 0.2 fm, the transverse radius
of the flux tube estimated by  \cite{Yanagihara:2018qqg}.
\end{enumerate}

Combining Eq.~(\ref{T44rrdegen}) with Eq.~(\ref{tracedegen}), we obtain
\begin {equation}
\left\langle T^R_{rr} (r) \right\rangle_{Q\bar{Q}}   <   - \left\langle T^R_{xx}(r)\right\rangle_{Q\bar{Q}} \;,
\label{trace_anomaly_2}
\end{equation}
which can be clearly seen from Fig.~3 of Ref.~\cite{Yanagihara:2018qqg},
 where the components
of  $\left\langle T_{ij}^R (r)\right\rangle_{Q \bar{Q}}$ were plotted.

The 'Maxwell' stress tensor does not include the contributions to Eq.~(\ref{Tmunua}) 
of the fluctuating color fields not measured in our simulations. 
Comparison of the spatial distributions
of the diagonal components of the Yang-Mills stress tensor with the corresponding
distributions of the 'Maxwell' stress tensor then provides a measure of the
contributions of the fluctuating color fields.

\section*{Acknowledgements}
This investigation was in part based on the MILC collaboration's public
lattice gauge theory code. See {\url{http://physics.utah.edu/~detar/milc.html}}.
Numerical calculations have been made possible through a CINECA-INFN
agreement, providing access to resources on MARCONI at CINECA.
AP, LC, PC, VC acknowledge support from INFN/NPQCD project.
FC acknowledges support from the DFG (Em\-my Noether Programme EN 1064/2-1).
VC acknowledges financial support from the INFN HPC{\_}HTC project.

\bibliography{qcd}

\end{document}